\documentclass[journal,onecolumn,12pt]{IEEEtran}%
\pdfoutput=1

\usepackage{ucs}
\usepackage[utf8x]{inputenc}
\usepackage{hyperref}
\usepackage{amsfonts}
\usepackage{amsmath}
\usepackage{amssymb}
\usepackage{graphicx}
\usepackage{amsthm}%

\providecommand{\U}[1]{\protect\rule{.1in}{.1in}}
\newtheorem{theorem}{Theorem}

\newtheorem{definition}{Definition}

\newtheorem{lemma}{Lemma}

\newtheorem{remark}[theorem]{Remark}

\newcommand{\Tr}{{\rm Tr}\;}

\newcommand{\nc}{\newcommand}
\nc{\rnc}{\renewcommand}
\nc{\beq}{\begin{equation}}
\nc{\eeq}{{\end{equation}}}
\nc{\beqa}{\begin{eqnarray}}
\nc{\eeqa}{\end{eqnarray}}
\nc{\lbar}[1]{\overline{#1}}
\nc{\bra}[1]{\langle#1|}
\nc{\ket}[1]{|#1\rangle}
\nc{\ketbra}[2]{|#1\rangle \langle#2|}
\nc{\braket}[2]{\langle#1|#2\rangle}
\nc{\proj}[1]{| #1\rangle \langle #1 |}
\nc{\avg}[1]{\langle#1\rangle}
\nc{\smfrac}[2]{\mbox{$\frac{#1}{#2}$}}
\nc{\tr}{\operatorname{tr}}
\nc{\tracedist}[1]{\Delta_{}\left( #1 \right)}
\nc{\fid}[1]{F\left( #1 \right)}

\nc{\ox}{\otimes}
\nc{\dg}{\dagger}
\nc{\dn}{\downarrow}
\nc{\cA}{{\cal A}}
\nc{\cB}{{\cal B}}
\nc{\cC}{{\cal C}}
\nc{\cD}{{\cal D}}
\nc{\cE}{{\mathcal E}}
\nc{\cF}{{\cal F}}
\nc{\cG}{{\cal G}}
\nc{\cH}{{\cal H}}
\nc{\cI}{{\cal I}}
\nc{\cJ}{{\cal J}}
\nc{\cK}{{\cal K}}
\nc{\cL}{{\cal L}}
\nc{\cM}{{\cal M}}
\nc{\cN}{{\cal N}}
\nc{\cO}{{\cal O}}
\nc{\cP}{{\cal P}}
\nc{\cR}{{\cal R}}
\nc{\cS}{{\cal S}}
\nc{\cT}{{\cal T}}
\nc{\cU}{{\cal U}}
\nc{\cX}{{\cal X}}
\nc{\cZ}{{\cal Z}}

\nc{\entI}{{\bf I}}
\nc{\entIarrow}{{\bf I}^{\leftarrow}}
\nc{\entH}{{\bf H}}
\nc{\entS}{{\bf S}}
\nc{\entHmin}{H_{\min}}

\nc{\entF}{{\bf E}_f}

\nc{\isom}{\simeq}

\nc{\rank}{\operatorname{rank}}
\nc{\rar}{\rightarrow}
\nc{\lrar}{\longrightarrow}
\nc{\polylog}{\operatorname{polylog}}
\nc{\poly}{\operatorname{poly}}
\nc{\1}{{\openone}}

\nc{\weight}{\textbf{w}}
\nc{\hamdist}{d_{H}}

\def\U{\Upsilon}

\nc{\Sp}{{{\mathbb S}}}
\nc{\RR}{{{\mathbb R}}}
\nc{\CC}{{{\mathbb C}}}
\nc{\FF}{{{\mathbb F}}}
\nc{\NN}{{{\mathbb N}}}
\nc{\ZZ}{{{\mathbb Z}}}
\nc{\PP}{{{\mathbb P}}}
\nc{\QQ}{{{\mathbb Q}}}
\nc{\UU}{{{\mathbb U}}}
\nc{\OO}{{{\mathbb O}}}
\nc{\EE}{{{\mathbb E}}}
\nc{\id}{{\operatorname{id}}}

\nc{\qubitchannel}{\id_2}
\nc{\bitchannel}{\overline{\id}_2}

\nc{\be}{\begin{equation}}
\nc{\ee}{\end{equation}}
\nc{\bea}{\begin{eqnarray}}
\nc{\eea}{\end{eqnarray}}
\nc{\<}{\langle}
\rnc{\>}{\rangle}
\nc{\Hom}[2]{\mbox{Hom}(\CC^{#1},\CC^{#2})}
\nc{\rU}{\mbox{U}}

\nc{\ob}[1]{#1}

\def\mcal{\mathcal}
\def\eps{\varepsilon}

\usepackage{ifthen}
\newboolean{THESIS}
\setboolean{THESIS}{true}

\newboolean{WITHAPDX}
\setboolean{WITHAPDX}{true}
\def\ifthen#1#2{\ifthenelse{#1}{#2}{} }

\def\PIavg{\Pi^n_{\bar{\rho},\delta}}
\def\PIone{\Pi_{u^n_1(\ell_1)}}

\def\rhoNulul{\rho_{\ell_1,\ell_2} }%

\def\omgu{\omega^{B_1}_{u_1}}

\def\AtypGua{\mcal{A}^{(m_a)}_{Z_{1p}|u_a,\delta}}
\def\AtypGuaun{\mcal{A}^{(m_a)}_{Z_{1p}|u_au^n_2,\delta}}

\def\ExpXW{\mathop{\mathbb{E}}_{\scriptscriptstyle X^{n},W^{n}}}
\def\ExpW{\mathop{\mathbb{E}}_{\scriptscriptstyle W^{n}}}
\def\ExpXGW{\mathop{\mathbb{E}}_{\scriptscriptstyle X^{n}|W^{n}}}

\def\PIrhoW{\Pi_{W^{n}(m_2)} }
\def\PIrhoWone{\Pi_{W^{n}(1)} }

\def\pullleft{\!\!\!\!\!\!}

\begin{document}

	\title{Classical codes for quantum broadcast channels}

\author{
	Ivan Savov and 
	Mark M. Wilde \thanks{I.S.~and M.M.W.~were with the School of Computer Science, 
	McGill University,  Montr\'eal, Qu\'ebec, Canada when conducting parts of this research. 
	I.S.~is now with Minireference Publishing. 
	M.M.W.~is now with the Hearne Institute for Theoretical Physics, the Department of Physics and Astronomy, 
	and the Center for Computation and Technology at Louisiana State University, Baton Rouge, LA 70803. 
	This work was presented in part at the 2012 IEEE International Symposium on Information Theory.}
}

	\maketitle
	
	\begin{abstract}
		We present two approaches for transmitting classical information
		over quantum broadcast channels. 
		The first technique is a quantum generalization of the superposition coding scheme
		for the classical broadcast channel.
		We use a quantum simultaneous nonunique decoder and obtain a proof of the rate region 
		stated in [Yard \textit{et al.}, IEEE Trans.~Inf.~Theory 57 (10), 2011].
		Our second result is a quantum generalization of the Marton coding scheme.
		The error analysis for the quantum Marton region makes use of ideas in our
		earlier work and an idea recently
		presented by Radhakrishnan \textit{et al.}~in arXiv:1410.3248.
		Both results exploit recent advances in quantum simultaneous decoding developed 
		in the context of quantum interference channels.		
	\end{abstract}

\section{Introduction}

	How can a broadcast station communicate separate messages to two 
	receivers using a single antenna?
	Two well known strategies \cite{el2010lecture} for transmitting information over broadcast channels are 
	superposition coding \cite{C72,B73} and Marton multicoding using correlated
	 auxiliary random variables \cite{M79}.
	In this paper, we prove that these strategies can be adapted to the quantum setting
	by constructing random codebooks and matching decoding measurements that
	 have asymptotically vanishing 
	 error in the limit of many uses of the channel.
	
	Sending classical data over a quantum channel is one of the 
	fundamental problems of quantum information theory~\cite{wilde2011book}.
	Single-letter formulas are known for classical-quantum point-to-point channels \cite{H98,SW97}
	and multiple access channels \cite{winter2001capacity}.
	Classical-quantum channels are a useful abstraction for studying general 
	quantum channels and correspond to the transmitters being restricted to classical encodings.
	Codes for classical-quantum channels (c-q channels), when augmented with an extra optimization 
	over the possible input states, directly generalize to codes for quantum channels.
	Furthermore, it is known that classical encoding (coherent-state encoding using classical Gaussian codebooks)
	is sufficient to achieve the capacity of phase-insensitive quantum Gaussian channels, which is a 
	realistic model for optical communication links \cite{PhysRevA.63.032312,PhysRevLett.92.027902,GHG13}.

	Previous work on quantum broadcast channels includes \cite{YHD2006,guha2007classical,DHL10}.
	Yard \textit{et al.}~consider both classical and quantum communication over quantum broadcast channels
	 and state a superposition coding inner bound in their Theorem~1 similar to that stated in our Theorem~\ref{thm:sup-coding-inner-bound} \cite{YHD2006}.
	 However, it is unclear to us  % COMMENT: this sounds weak... why is it not clear to us?
	 whether the proof given for their Theorem~1 is complete (we elaborate on this point in what follows).
	 Relying on Theorem~1 of \cite{YHD2006}, Ref.~\cite{guha2007classical} discusses classical communication over a
				 bosonic broadcast channel. 
	 Ref.~\cite{DHL10} establishes a Marton rate region for quantum communication.

	In this paper, we derive  
	two achievable rate regions 
	for classical-quantum broadcast channels
	by exploiting error analysis techniques developed
	in the context of quantum interference channels \cite{FHSSW11,S11a}.
	In Section~\ref{sec:superposition-coding}, we prove achievability of the 
	superposition coding inner bound (Theorem~\ref{thm:sup-coding-inner-bound}),
	by using a quantum simultaneous nonunique decoder at one of the receivers.
	In Section~\ref{sec:marton} we prove that  the quantum Marton rate region with no common message is achievable
	(Theorem~\ref{thm:marton-no-common}).
	In the Marton coding scheme, the sub-channels to each receiver are essentially point-to-point,
	but it turns out that two techniques which we call the ``projector trick'' and
	``overcounting'' \cite{RSW14} seem to be 
	necessary in the proof.
	We discuss open problems and give an outlook for the future %
	in Section~\ref{sec:conclusion}.
	
	\textit{Note}: The original justification for the quantum Marton region given in our earlier work \cite{SW12ISIT}
	contained a gap, which was identified by Pranab Sen and relayed to us by Andreas Winter.
	This gap was addressed
    in the related paper \cite{RSW14}, where  an achievable region in the
    ``one-shot'' Marton coding setting was established.
	Here we show how to apply the overcounting method in order to close
	the aforementioned gap in our earlier work.

\section{Preliminaries}

	\subsubsection{Notation}

		We denote classical random variables as $X,U,W$, whose realizations are elements
		of the respective finite alphabets $\mcal{X}, \mcal{U}, \mcal{W}$.
		Let $p_X$, $p_U$, $p_W$ denote their corresponding probability distributions.
		We denote quantum systems as $A$, $B$, and $C$ and their corresponding Hilbert
		spaces as $\mathcal{H}^{A}$, $\mathcal{H}^{B}$, and $\mathcal{H}^{C}$.
		We represent quantum states of a system~$A$ with a density operator $\rho^{A}$,
		which is a positive semi-definite operator with unit trace.
		Let $H(A)_{\rho}\equiv-\text{Tr}\left\{  \rho^{A}\log\rho^{A}\right\}$ 
		denote the von Neumann entropy of the state $\rho^{A}$. 
		A classical-quantum channel, $\mcal{N}^{X\to B}$, 
		is represented by the set of $|\mcal{X}|$ possible output states 
		$\{ \rho^B_x \equiv \mcal{N}^{X\to B}(x) \}$, meaning that a classical 
		input of $x$ leads to a quantum output $\rho^B_x$.
		In a communication scenario, the decoding operations performed by the 
		receivers correspond to quantum measurements on the outputs of the channel.
		A quantum measurement %
		is a positive operator-valued measure (POVM) 
		$\left\{ \Lambda_{m}\right\}_{m\in\left\{  1,\ldots,|\mathcal{M}|\right\}  }$ on
		the system $B^n$, the output of which we denote $M^{\prime}$. 
		To be a valid POVM, the set of $|\mcal{M}|$ operators $\Lambda_{m}$ must all be positive semi-definite
		and sum to the identity:  $\Lambda_{m} \geq 0, \,\,\, \sum_{m}\Lambda_{m}=I$.

	\subsubsection{Definitions and useful lemmas}

		We define a classical-quantum-quantum broadcast channel as the following map:%
		\begin{equation}
		x\rightarrow\rho_{x}^{B_{1}B_{2}},\label{eq:cqq-broadcast}%
		\end{equation}
		where $x$ is a classical letter in an alphabet $\mathcal{X}$ and 
		$\rho_{x}^{B_{1}B_{2}}$ is a density operator on the tensor product Hilbert
		space for systems $B_{1}$ and $B_{2}$. The model is such that when the sender
		inputs a classical letter $x$, Receiver~1 obtains system $B_{1}$, and
		Receiver~2 obtains system $B_{2}$.
		Since Receiver~1 does not have access to the $B_2$ part of the state $\rho_{x}^{B_{1}B_{2}}$,
		we model his state as $\rho_{x}^{B_{1}} = \text{Tr}_{B_2}\left[  \rho_{x}^{B_{1}B_{2}}  \right]$,
		where $\text{Tr}_{B_2}$ denotes the partial trace over Receiver~2's system.
		
		\begin{lemma}[Gentle Operator Lemma for Ensembles \cite{itit1999winter}] \label{lem:gentle-operator}
		Given an ensemble $\left\{  p_{X}\left(  x\right)  ,\rho_{x}\right\}  $ with
		expected density operator $\rho\equiv\sum_{x}p_{X}\left(  x\right)  \rho_{x}$,
		suppose that an operator $\Lambda$ such that $I\geq\Lambda\geq0$ succeeds
		with high probability on the state $\rho$:%
		\be
		\operatorname{Tr}\left\{  \Lambda\rho\right\}  \geq1-\varepsilon.
		\ee
		Then the subnormalized state $\sqrt{\Lambda}\rho_{x}\sqrt{\Lambda}$ is close
		in expected trace distance to the original state $\rho_{x}$:%
		\be
		\mathbb{E}_{X}\left\{  \left\Vert \sqrt{\Lambda}\rho_{X}\sqrt{\Lambda}%
		-\rho_{X}\right\Vert _{1}\right\}  \leq2\sqrt{\varepsilon}.
		\ee

		\end{lemma}
		
		\noindent
		The following lemma appears in  \cite[Lemma 2]{hayashi2003general}. When using it for the
		square-root measurement in \eqref{eq:square-root-POVM-generic}, we choose
		$S = \Pi'_m$ and $T = \sum_{k\neq m} \Pi'_k$.
		\begin{lemma}[Hayashi-Nagaoka] \label{lem:HN-inequality}
		The Hayashi-Nagaoka operator inequality applies to a positive operator $T$ and
an operator $S$ where $0\leq S\leq I$:%
\be
I-\left(  S+T\right)  ^{-\frac{1}{2}}S\left(  S+T\right)  ^{-\frac{1}{2}}%
\leq2\left(  I-S\right)  +4T.
\ee
		\end{lemma}

	\subsubsection{Information processing task}

		The task of communication over a broadcast channel 
		is to use $n$ independent instances of the channel in order to communicate with Receiver~1 at a rate $R_1$
		and to Receiver~2 at a rate $R_2$.
		More specifically, the sender chooses a pair of messages $(m_1,m_2)$ from message sets
		$\mathcal{M}_i\equiv \left\{  1,2,\ldots,|\mathcal{M}_i|\right\} $, where $|\mathcal{M}_i|=2^{nR_{i}}$,
		and encodes these messages into an $n$-symbol codeword $x^{n}\left( m_1,m_2\right)\in \mathcal{X}^n$
		suitable as input for the $n$ channel uses.

		The output of the channel is a quantum state %
		of the form:
		\begin{align}
			\mathcal{N}^{\otimes n}\left( x^{n}(m_1,m_2) \right)
			& \equiv			
			\rho_{x^{n}\left(  m_1,m_2\right)  }^{B_{1}^{n}B_{2}^{n}} . 
		\end{align}
		  where
		  $
		  \rho_{x^{n}}^{B_{1}^{n}B_{2}^{n}} \equiv \rho_{x_1}^{B_{11}B_{21}} \otimes
		  \cdots \otimes \rho_{x_n}^{B_{1n}B_{2n}}.
		  $
		To decode the message $m_1$ intended for him,
		Receiver~1 performs a POVM
		$\left\{ \Lambda_{m_1}\right\}  _{m_1\in\left\{  1,\ldots,|\mathcal{M}_1|\right\}  }$ on
		the system $B_{1}^n$, the output of which we denote $M^{\prime}_1$. 
		Receiver~2 similarly performs a POVM 
		$\left\{  \Gamma_{m_2}\right\} _{m_2\in\left\{  1,\ldots ,|\mathcal{M}_2|\right\}  }$
		on the system  $B_{2}^n$,
		and the random variable associated with the outcome is denoted $M^{\prime}_2$.
		
		An error occurs whenever either of the receivers decodes the message incorrectly.
		The probability of error for a particular message pair $(m_1,m_2)$ is
		\begin{align}
			p_{e}\left(  m_1,m_2\right)   
				\equiv
				\text{Tr}
				\left\{  
					\left(  I-\Lambda_{m_1}\otimes\Gamma_{m_2}\right)
					\rho_{x^{n}\left(  m_1,m_2\right)  }^{B_{1}^{n}B_{2}^{n}} 
				\right\},
		\end{align}
		where the operator $\left(  I-\Lambda_{m_1}\otimes\Gamma_{m_2}\right)$ represents
		the complement of the correct decoding outcome.

		    \begin{definition}%
			An $(n,R_1,R_2,\varepsilon)$ broadcast channel code  consists
			of a codebook
			$\{x^n(m_1,m_2)\}_{m_1\in \mathcal{M}_1, m_2\in \mathcal{M}_2}$
			and two decoding POVMs  
			$\left\{ \Lambda_{m_1}\right\}_{m_1\in \mathcal{M}_1}$ 
			and 
			$\left\{  \Gamma_{m_2}\right\}_{m_2\in \mathcal{M}_2}$
			such that the average probability of error 
			$\overline{p}_{e}$ is bounded from above as
			\begin{align}
				\overline{p}_{e}  
				&  \equiv 
					\frac{1}{|\mathcal{M}_1||\mathcal{M}_2|}\sum_{m_1,m_2}p_{e}\left(  m_1,m_2\right) %
				 \leq \varepsilon.
			\end{align}

		    \end{definition}  
		
		A rate pair $\left(  R_{1},R_{2}\right)  $ is \textit{achievable} if there exists
		an $\left(  n,R_{1}-\delta,R_{2}-\delta,\varepsilon\right)  $ quantum broadcast channel code
		for all $\varepsilon,\delta>0$ and sufficiently large $n$.

		When devising coding strategies for c-q channels, 
		the main obstacle to overcome 
		is the construction of a decoding POVM that correctly
		decodes the messages.
		Given a set of positive operators $\{ \Pi^\prime_m \}$ which are suitable for detecting each message,
		we can construct a POVM by \emph{normalizing} them using the square-root
		  measurement \cite{H98,SW97}:
		\begin{align}
		\Lambda_{m} &  \equiv 
		\left(  
			\sum_{k}\Pi_{k}^{\prime}
		\right)^{-\frac{1}{2}}
		 \Pi_{m}^{\prime}
		\left(
			\sum_{k}\Pi_{k}^{\prime}
		\right)^{-\frac{1}{2}}. \label{eq:square-root-POVM-generic} 
		\end{align}
		Thus, the search for a decoding POVM is reduced to the problem of finding
		positive operators $\Pi^\prime_m$ apt at detecting and distinguishing  
		the output states produced by each of the possible input messages
		($\text{Tr}\left[\Pi^\prime_m\ \rho_m\right] \geq 1- \varepsilon'$ and 
		$\text{Tr}\left[\Pi^\prime_{m}\ \rho_{m^\prime\neq m}\right] \leq \varepsilon'$ for some small $\varepsilon'>0$).

\section{Superposition coding inner bound}	\label{sec:superposition-coding}

		One possible strategy for the broadcast channel is to send a message
		at a rate that is low enough so that both receivers are able to decode.
		Furthermore, if we assume that Receiver~1 has a better reception signal, 
		then the sender can encode a further message \emph{superimposed} on top of the common message
		that Receiver 1 will be able to decode \emph{given} the common message.
		The sender encodes the common message at rate $R_2$ using a codebook
		generated from a probability distribution $p_W(w)$, 
		and the additional message for Receiver~1 at rate $R_1$ using a conditional
		codebook with distribution $p_{X|W}(x|w)$.

		\begin{theorem}[Superposition coding inner bound] 	\label{thm:sup-coding-inner-bound}
		A rate pair $\left(  R_{1},R_{2}\right)  $
		is achievable for the quantum broadcast channel in \eqref{eq:cqq-broadcast} if
		it satisfies the following inequalities:%
		\begin{align}
			R_1 & \leq I(X; B_1 | W)_\theta, \\
			R_2  &\leq I(W; B_2)_\theta, \\
			R_1 + R_2  &\leq I(X;B_1)_\theta,
		\end{align}
		where the above information quantities are with respect to a state $\theta^{WXB_1B_2} $ of the form
		\be
			\sum_{w,x} 
				p_W(w)p_{X|W}(x|w) \
				\ketbra{w}{w}^{W}
				\otimes
				\ketbra{x}{x}^{X}
				\otimes
				\rho_x^{B_1B_2}.\label{eq:code-state}
		\ee		
It suffices to take the cardinality of the alphabet $\mathcal{W}$ for $W$ to be no larger than
$\min\{|\mathcal{X}|, |B_1|^2 + |B_2|^2-1\}$, where
$\mathcal{X}$ is the input alphabet of the channel.
		\end{theorem}

		\begin{IEEEproof}
		The idea of the proof given below is to exploit superposition encoding
		and a quantum simultaneous nonunique decoder for the decoding of
		the first receiver \cite{C72,B73}.
		We use a standard HSW\ decoder for the second receiver \cite{H98,SW97}. The cardinality bound follows directly from Appendix~A of \cite{YHD2006}.

	\textbf{Codebook generation.} The sender randomly and independently generates
		$M_{2}$ sequences $w^{n}\!\left(  m_{2}\right)  $ according to the product
		distribution
		\be
		p_{W^{n}}\!\left(  w^{n}\right)  \equiv\prod\limits_{i=1}^{n}p_{W}\!\left(w_{i}\right).
		\ee
		For each sequence $w^{n}\!\left(  m_{2}\right)  $, the sender then randomly and
		conditionally independently generates $M_{1}$ sequences $x^{n}\!\left(
		m_{1},m_{2}\right)  $ according to the product distribution: 
		\be
		p_{X^{n}|W^{n}}\!\left(  x^{n}|w^{n}\!\left(  m_{2}\right)  \right)  \equiv
		\prod\limits_{i=1}^{n}p_{X|W}\!\left(  x_{i}|w_{i}\!\left(  m_{2}\right)  \right).
		\ee
		The sender then transmits the codeword $x^{n}\!\left(  m_{1},m_{2}\right)  $ if
		she wishes to send $\left(  m_{1},m_{2}\right)  $.

	\textbf{POVM\ Construction}. We now describe the POVMs that the
		receivers employ in order to decode the transmitted messages. 
		First consider the state we obtain from (\ref{eq:code-state}) by tracing over
		the $B_{2}$ system:%
		\be
		\rho^{WXB_{1}}=\sum_{w,x}p_{W}\!\left(  w\right)  \ p_{X|W}\!\left(  x|w\right)
		\ \vert w\rangle\langle w\vert ^{W}\otimes\vert
		x\rangle\langle x\vert ^{X}\otimes\rho_{x}^{B_{1}}.
		\ee
		Further tracing over the $X$ system gives%
		\begin{align}
		\rho^{WB_{1}}  &  =\sum_{w}p_{W}\!\left(  w\right)  \ \vert w\rangle
		\langle w\vert ^{W}\otimes\sigma_{w}^{B_{1}},
		\end{align}
		where $\sigma_{w}^{B_{1}}\equiv \sum_{x}p_{X|W}\!\left(  x|w\right)
		\rho_{x}^{B_{1}}$.
		For the first receiver, we exploit a square-root decoding POVM\ as in \eqref{eq:square-root-POVM-generic} 
		based on the following positive operators:
		\be
		\Pi_{m_{1},m_{2}}^{\prime}\equiv\Pi\ \PIrhoW %
		\ \Pi_{X^{n}\!\left(  m_{1},m_{2}\right)  }\ \PIrhoW \ \Pi, \label{eq:rec-1-POVM}
		\ee
		where we have made the abbreviations%
		\begin{align}
		\Pi   \equiv\Pi_{\rho,\delta}^{B_{1}^{n}}  , \,\,\,\,\,\,\,\,\,
		\PIrhoW     \equiv\Pi_{\sigma_{W^{n}\!\left(
		m_{2}\right)  },\delta}^{B_{1}^{n}}, \,\,\,\,\,\,\,\,\,
		\Pi_{X^{n}\!\left(  m_{1},m_{2}\right)  }    \equiv\Pi_{\rho_{X^{n}\!\left(
		m_{1},m_{2}\right)  },\delta}^{B_{1}^{n}}.
		\end{align}
		The above projectors are weakly typical projectors \cite[Section 14.2.1]{wilde2011book}
		  defined with respect to the states $\rho^{\otimes n}$, $\sigma^{B_{1}^{n}}_{W^{n}\!\left(
		m_{2}\right)}$, and $\rho^{B_{1}^{n}}_{X^{n}\!\left(
		m_{1},m_{2}\right)}$.
		
		Consider now the state in (\ref{eq:code-state}) as it looks from the point of view 
		of Receiver~2. 
		If we trace over the $X$ and $B_{1}$
		systems, we obtain the following state:%
		\begin{align}
		\rho^{WB_{2}}  =\sum_{w}p_{W}\!\left(  w\right)  \ \vert w\rangle
		\langle w\vert ^{W}\otimes\sigma_{w}^{B_{2}},
		\end{align}
		where $\sigma_{w}^{B_{2}}\equiv \sum_{x}p_{X|W}\!\left(
		x|w\right)  \rho_{x}^{B_{2}}$.
		For the second receiver, we exploit a standard HSW\ decoding POVM\ that is with
		respect to the above state---it is a square-root
		measurement as in \eqref{eq:square-root-POVM-generic},
		based on the following positive operators:
		\be
			\Pi_{m_2}^{\prime B_{2}^{n} }
			=\Pi_{\rho,\delta}^{B_{2}^{n} }\ \Pi_{\sigma_{W^{n}(m_2)},\delta}^{B_{2}}\ \Pi_{\rho,\delta}^{B_{2}^{n}}, \label{eq:rec-2-POVM}
		\ee
		where the above projectors are weakly typical projectors defined with respect to
		  $\rho^{\otimes n}$ %
		  and $\sigma^{B_{2}^{n}}_{W^{n}\!\left(
		m_{2}\right)}$.

	\textbf{Error analysis}.		
		We now analyze the expectation of the average error probability for the first
		receiver with the POVM defined by (\ref{eq:square-root-POVM-generic}) and (\ref{eq:rec-1-POVM}):
		\begin{multline}
		\ExpXW \left\{  \frac{1}{M_{1}M_{2}}\sum_{m_{1},m_{2}%
		}\text{Tr}\left\{  \left(  I-\Gamma_{m_{1},m_{2}}^{B_{1}^{n}}\right)
		\rho_{X^{n}\!\left(  m_{1},m_{2}\right)  }^{B_{1}}\right\}  \right\}  \\
		=\frac{1}{M_{1}M_{2}}\sum_{m_{1},m_{2}}\ExpXW \left\{
		\text{Tr}\left\{  \left(  I-\Gamma_{m_{1},m_{2}}^{B_{1}^{n}}\right)
		\rho_{X^{n}\!\left(  m_{1},m_{2}\right)  }^{B_{1}}\right\}  \right\}.
		\end{multline}
		Due to the above exchange between the expectation and the average and the
		symmetry of the code construction (each codeword is selected randomly and
		independently), it suffices to analyze the expectation of the average error
		probability for the first message pair $\left(  m_{1}=1,m_{2}=1\right)  $,
		i.e., the last line above is equal to
		\be
		  \ExpXW \left\{  \text{Tr}\left\{  \left(  I-\Gamma
		_{1,1}^{B_{1}^{n}}\right)  \rho_{X^{n}(1,1)  }^{B_{1}}\right\}
		\right\}  .
		\ee
		Using the Hayashi-Nagaoka operator inequality (Lemma~\ref{lem:HN-inequality} in the appendix),
		we obtain the following upper bound on this term:
		\be
		2  \ExpXW \left\{  \text{Tr}\left\{  \left(
		I-\Pi_{1,1}^{\prime}\right)  \rho_{X^{n}(1,1)  }^{B_{1}}\right\}
		\right\}  \label{eq:after-HN}
		+4  \sum_{\left( m_{1},m_{2}\right)  \neq\left(  1,1\right)  }
		\ExpXW \left\{  \text{Tr}\left\{  \Pi_{m_{1},m_{2}}^{\prime}
		\rho_{X^{n}(1,1)  }^{B_{1}}\right\}  \right\}. 
		\ee
		
		We begin by bounding the first term above. Consider the following
		chain of inequalities:%
		\begin{align}
		 \ExpXW \left\{  \text{Tr}\left\{  \Pi_{1,1}^{\prime}
		\rho_{X^{n}(1,1)  }^{B_{1}}\right\}  \right\}   
		& =
		\ExpXW
		\left\{  \text{Tr}\left\{  \Pi\ \PIrhoWone
		\Pi_{X^{n}(1,1)  }\ \PIrhoWone\ \Pi
		\ \rho_{X^{n}(1,1)  }^{B_{1}}\right\}  \right\}  \\
		&  \geq\ExpXW \left\{  \text{Tr}\left\{  \Pi_{X^{n}\!\left(
		1,1\right)  }\rho_{X^{n}(1,1)  }^{B_{1}}\right\}  \right\}
		-\ExpXW \left\{  \left\Vert \rho_{X^{n}(1,1)  }^{B_{1}}-\Pi\ \rho_{X^{n}(1,1)  }^{B_{1}}%
		\ \Pi\right\Vert _{1}\right\}  \nonumber \\
		&  \ \ \ \ \ -\ExpXW \left\{  \left\Vert \rho_{X^{n}(1,1)  }^{B_{1}}
		-\PIrhoWone\ \rho_{X^{n}(1,1)  }^{B_{1}}\ \PIrhoWone\right\Vert _{1}\right\}
		\\
		&  \geq1-\varepsilon-4\sqrt{\varepsilon},
		\end{align}
		where the first inequality follows from the inequality%
		\begin{equation}
		\text{Tr}\left\{  \Lambda\rho\right\}  \leq\text{Tr}\left\{  \Lambda
		\sigma\right\}  +\left\Vert \rho-\sigma\right\Vert _{1}, \label{eq:trace-inequality}
		\end{equation}
		which holds for all subnormalized states $\rho$ and $\sigma$, and $\Lambda$ such that $0\leq
		\Lambda\leq I$. 
		The second inequality follows from the Gentle Operator Lemma for 
		ensembles (see Lemma~\ref{lem:gentle-operator} in the appendix) and the properties of typical projectors
		  for sufficiently large $n$.

		We now focus on bounding the second term in (\ref{eq:after-HN}). We can expand this term
		as follows:%
		\be
		\sum_{m_{1}\neq1}\ExpXW \left\{  \text{Tr}\left\{  \Pi_{m_{1},1}
		^{\prime}\ \rho_{X^{n}(1,1)  }^{B_{1}}\right\}  \right\}
		+\sum_{\substack{m_{1}, \\ m_{2} \neq1}}\ExpXW \left\{
		\text{Tr}\left\{  \Pi_{m_{1},m_{2}}^{\prime}\ \rho_{X^{n}(1,1)
		}^{B_{1}}\right\}  \right\}.
		\label{eq:error-terms} 
		\ee

		Consider the first term in (\ref{eq:error-terms}):%
		\begin{align}
		& \pullleft \pullleft  \sum_{m_{1}\neq1}
		\ExpXW \left\{  \text{Tr}\left\{
		\Pi_{m_{1},1}^{\prime}\ \rho_{X^{n}(1,1)  }^{B_{1}}\right\}
		\right\}  \\
		&  =\sum_{m_{1}\neq1}\ExpXW   \text{Tr}\left\{
		\Pi\ \PIrhoWone\ \Pi_{X^{n}\!\left(  m_{1},1\right)  }
		\ \PIrhoWone\ \Pi\ \rho_{X^{n}(1,1)  }^{B_{1}}\right\}    \\
		& \leq
		2^{n\left[  H\!\left(  B_{1}|WX\right)  +\delta\right]  }  \sum_{m_{1}\neq1}
		\ExpXW 
		\left\{  \text{Tr}\left[  \Pi\ \PIrhoWone 
		\rho_{X^{n}\!\left(  m_{1},1\right)  }
		\PIrhoWone 
		\Pi\ \rho_{X^{n}(1,1)  }^{B_{1}}\right]  \right\}  \\
		&  =
		2^{n\left[  H\!\left(  B_{1}|WX\right)  +\delta\right]  }
		\sum_{m_{1}\neq1}
		\ExpW
		\bigg\{  \text{Tr}[   
		\PIrhoWone
		\ExpXGW \left\{  \rho_{X^{n}\left(m_{1},1\right)  }\right\}  \ 
		\PIrhoWone\   \Pi\ \ExpXGW 
		\left\{  \rho_{X^{n}\left(  1,1\right)  }^{B_{1}}\right\} 
		\Pi\ ] \bigg\} \\
		&  =
		2^{n\left[  H\left(  B_{1}|WX\right)  +\delta\right]  } 
		\sum_{m_{1}\neq 1} 
		\ExpW 
		\left\{  \text{Tr}\left\{  
			\Pi\ 
			\PIrhoWone  \sigma_{W^{n}\left(  1\right)  }\ \PIrhoWone\ 
			\Pi\ \sigma_{W^{n}\left(  1\right)  }
		\right\}  \right\}  \\
		&  \leq
		2^{n\left[  H\left(  B_{1}|WX\right)  +\delta\right]  } \,2^{-n\left[ H\left(  B_{1}|W\right)  -\delta\right]  }
		\sum_{m_{1}\neq1} 
		\ExpW 
		\left\{  \text{Tr}\left\{  \Pi\ 
		\PIrhoWone\ \Pi\ \sigma_{W^{n}\left(  1\right)  }\right\}  \right\}  \\
		&  \leq
		2^{n\left[  H\left(  B_{1}|WX\right)  +\delta\right]  }\,
		2^{-n\left[H\left(  B_{1}|W\right)  -\delta\right]  }\ 
		\sum_{m_{1}\neq1}
		\ExpW 
		\left\{  \text{Tr}\left\{  \sigma_{W^{n}\left(  1\right)  }\right\}
		\right\}  \\
		&  \leq
		2^{-n\left[  I\left(  X;B_{1}|W\right)  -2\delta\right]  }\ M_{1}.
		\end{align}
		The first inequality is due to the \emph{projector trick} inequality \cite{GLM10,S11a,FHSSW11}
		which states that
		\begin{align}
			\Pi_{X^{n}\left(  m_{1},1\right)  }
		& \leq
			2^{n\left[  H\left(  B_{1}|WX\right)  +\delta\right]  }\,\rho_{X^{n}\left( m_1,1\right)  }^{B_{1}}. \label{eq:projector-trick}
		\end{align} Note that this inequality is a straightforward consequence of the following standard typicality operator inequality and the fact that $\Pi_{X^{n}\left(  m_{1},1\right)  }$ and 
		$\rho_{X^{n}\left( m_1,1\right)  }^{B_{1}}$ commute:
		\begin{equation}
2^{-n\left[  H\left(  B_{1}|WX\right)  +\delta\right]  }
		\Pi_{X^{n}\left(  m_{1},1\right)  }
		 \leq
		\Pi_{X^{n}\left(  m_{1},1\right)  } \,
			\rho_{X^{n}\left( m_1,1\right)  }^{B_{1}} \,
			\Pi_{X^{n}\left(  m_{1},1\right)  }.
		\end{equation}
		The second inequality follows from the properties of typical projectors:
		\be\PIrhoWone  \sigma_{W^{n}\left(  1\right)  }\ \PIrhoWone \leq 2^{-n\left[H\left(  B_{1}|W\right)  -\delta\right]  }\PIrhoWone.\ee

		Now consider the second term in (\ref{eq:error-terms}):%
		\begin{align}
		&  \pullleft \pullleft
		\sum_{\substack{m_{1}, \\ m_{2} \neq1}}
		\ExpXW 
		\left\{  \text{Tr}\left\{
		\Pi_{m_{1},m_{2}}^{\prime}\ \rho_{X^{n}\left(  1,1\right)  }^{B_{1}}\right\}
		\right\}  \\
		&  = 
		\sum_{\substack{m_{1}, \\ m_{2} \neq1}} 
		\ExpXW 
		\left\{  \text{Tr}
		\left[  \Pi \Pi_{W^{n}(m_2)  } \Pi_{X^{n}\left(  m_{1},m_{2}\right)  } \PIrhoW \Pi\ \rho_{X^{n}\left(1,1\right)  }^{B_{1}}\right]  \right\}  \\
		&  = 
		\sum_{\substack{m_{1}, \\ m_{2} \neq1}}
		\text{Tr}\left[  \ExpXW 
		\bigg\{  \PIrhoW \ \Pi_{X^{n}\left(  m_{1},m_{2}\right)  }\ \PIrhoW \right\}  \Pi 
		\ExpXW \left\{  \rho_{X^{n}\left(  1,1\right) }^{B_{1} } \right\}  \Pi\  \bigg]  \\
		&  = 
		\sum_{\substack{m_{1}, \\ m_{2} \neq1}} 
		\text{Tr}\left\{  \ExpXW \left\{  \PIrhoW  \Pi_{X^{n}\left(  m_{1},m_{2}\right)  } \PIrhoW \right\}  
		\ \Pi  \rho^{\otimes n} \Pi  \right\} \\
		&  \leq
		2^{-n\left[  H\left(  B_{1}\right)  -\delta\right]  } 
		\sum_{\substack{m_{1}, \\ m_{2} \neq1}} 
		\text{Tr}\left[  \ExpXW  \left\{  
			\PIrhoW  \Pi_{X^{n}\left(  m_{1},m_{2}\right)  }
			\PIrhoW \right\}  \Pi 
		\right]  \\
		&  =
		2^{-n\left[  H\left(  B_{1}\right)  -\delta\right]  } 
		\sum_{\substack{m_{1}, \\ m_{2} \neq1}} 
		\ExpXW    \text{Tr}\left[  
		\Pi_{X^{n}\left(  m_{1},m_{2}\right)  } \PIrhoW 
		\Pi \PIrhoW \right]   \\
		&  \leq
		2^{-n\left[  H\left(  B_{1}\right)  -\delta\right]  }
		\sum_{m_{2}	\neq1,\ m_{1}}\ExpXW 
		\left\{  \text{Tr}\left\{  \Pi_{X^{n}\left(  m_{1},m_{2}\right)  }\right\}  \right\}  \\
		&  \leq
		2^{-n\left[  H\left(  B_{1}\right)  -\delta\right]  }\ 
		2^{n\left[	H\left(  B_{1}|WX\right)  +\delta\right]  }\ 
		M_{1}\ M_{2}\\
		&  =
		2^{-n\left[  I\left(  WX;B_{1}\right)  -2\delta\right]  }\ M_{1}\ M_{2}\\
		& =
		2^{-n\left[  I\left(  X;B_{1}\right)  -2\delta\right]  }\ M_{1}\ M_{2}.
		\end{align}
		The equality $I(WX;B_1)=I(X;B_1)$ follows from
		the way the codebook is constructed (i.e., the Markov chain $W-X-B$),
		  as discussed also in \cite{S11a}.
		This completes the error analysis for the first receiver.
		
		 		For the second receiver, the decoding error analysis 
		follows from the HSW\ coding theorem. We now present this for completeness and tie
		  the coding theorem together so that the sender and two receivers can agree on
		  a strategy that has asymptotically vanishing error probability in the large $n$ limit.
		The following bound holds for the 
		expectation of the average error probability for the second receiver if $n$ is
		sufficiently large:%
		\begin{align}
		&  \pullleft \pullleft \ExpXW \left\{  \frac{1}{M_{2}}\sum_{m_{2}}\text{Tr}%
		\left\{  \left(  I-\Lambda_{m_{2}}^{B_{2}^{n}}\right)  \rho_{X^{n}\left(
		m_{1},m_{2}\right)  }^{B_{2}^{n}}\right\}  \right\}  \\
		&  =\ExpW \left\{  \frac{1}{M_{2}}\sum_{m_{2}}\text{Tr}\left\{
		\left(  I-\Lambda_{m_{2}}^{B_{2}^{n}}\right)  \ExpXGW \left\{
		\rho_{X^{n}\left(  m_{1},m_{2}\right)  }^{B_{2}^{n}}\right\}  \right\}
		\right\} \\
		&  =\ExpW \left\{  \frac{1}{M_{2}}\sum_{m_{2}}\text{Tr}\left\{
		\left(  I-\Lambda_{m_{2}}^{B_{2}^{n}}\right)  \sigma_{W^{n}\left(
		m_{2}\right)  }^{B_{2}^{n}}\right\}  \right\}  \\
		&  \leq2\left(  \varepsilon+2\sqrt{\varepsilon}\right)  +4\ \left[  2^{-n\left[
		I\left( W;B_{2}\right)  -2\delta\right]  }\ M_{2}\right]  ,
		\end{align}
		where the last line follows from an analysis similar to that given above. 

		Putting everything together, the joint POVM\ performed by both receivers is of
		the form:
		\be
		\Gamma_{m_{1},m_{2}}^{B_{1}^{n}}\otimes\Lambda_{m_{2}'}^{B_{2}^{n}},
		\ee
		and the expectation of the average error probability for both receivers is
		bounded from above as%
		\begin{align}
		&\pullleft \pullleft  \ExpXW   \frac{1}{M_{1}M_{2}}\sum_{m_{1},m_{2}%
		}\text{Tr}\left\{  \left(  I-\Gamma_{m_{1},m_{2}}^{B_{1}^{n}}\otimes
		\Lambda_{m_{2}}^{B_{2}^{n}}\right)  \rho_{X^{n}\left(  m_{1},m_{2}\right)
		}^{B_{1}^{n}B_{2}^{n}}\right\}   \\
		&  \leq\ExpXW \left\{  \frac{1}{M_{1}M_{2}}\sum_{m_{1},m_{2}%
		}\text{Tr}\left\{  \left(  I-\Gamma_{m_{1},m_{2}}^{B_{1}^{n}}\right)
		\rho_{X^{n}\left(  m_{1},m_{2}\right)  }^{B_{1}^{n}}\right\}  \right\} \\
		&  \ \ \ \ \ \ +\ExpXW \left\{  \frac{1}{M_{1}M_{2}}%
		\sum_{m_{1},m_{2}}\text{Tr}\left\{  \left(  I-\Lambda_{m_{2}}^{B_{2}^{n}%
		}\right)  \rho_{X^{n}\left(  m_{1},m_{2}\right)  }^{B_{2}^{n}}\right\}
		\right\} \\
		&  \leq 4  \varepsilon+12\sqrt{\varepsilon}  + 4\ \left[
		2^{-n\left[  I\left(  W;B_{2}\right)  -2\delta\right]  }\ M_{2}\right] \\
		&   \qquad + 4\ \left[  2^{-n\left[
		I\left(  X;B_{1}|W\right)  -2\delta\right]  }\ M_{1}+2^{-n\left[  I\left(
		X;B_{1}\right)  -2\delta\right]  }\ M_{1}\ M_{2}\right] ,
		\end{align}
		where the first inequality follows from the following ``union bound'' operator inequality:%
		\be
		I^{B_{1}^{n}B_{2}^{n}}-\Gamma_{m_{1},m_{2}}^{B_{1}^{n}}\otimes\Lambda_{m_{2}%
		}^{B_{2}^{n}} 
		\leq\left(  I^{B_{1}^{n}B_{2}^{n}}-\Gamma_{m_{1},m_{2}}%
		^{B_{1}^{n}}\otimes I^{B_{2}^{n}}\right)  +\left(  I^{B_{1}^{n}B_{2}^{n}%
		}-I^{B_{1}^{n}}\otimes\Lambda_{m_{2}}^{B_{2}^{n}}\right)  , \label{eq:commuting-union-bnd}
		\ee
		and the second inequality follows from our previous estimates. Thus, as long
		as the sender chooses the message sizes $M_{1}$ and $M_{2}$ such that
		$M_{1}    \leq2^{n\left[  I\left(  X;B_{1}|W\right)  -3\delta\right]  }$,
		$M_{2}  \leq2^{n\left[  I\left(  W;B_{2}\right)  -3\delta\right]  }$,
		and $M_{1} M_{2}    \leq2^{n\left[  I\left(  X;B_{1}\right)  -3\delta\right]  }$,
		then there exists a particular code with asymptotically vanishing average
		error probability in the large $n$ limit.
		\end{IEEEproof}

\begin{remark}
It is unclear to us whether the proof of \cite[Theorem~1]{YHD2006} is complete. These authors begin their
proof by claiming that the region in Theorem~\ref{thm:sup-coding-inner-bound} is equivalent to the following region:
		\begin{align}
			R_1 & \leq I(X; B_1 | W)_\theta, \\
			R_2  &\leq I(W; B_2)_\theta, \\
			R_2  &\leq I(W;B_1)_\theta.
		\end{align}
The regions certainly intersect at the corner point associated with their successive decoding strategy,
but the full regions for a fixed distribution do not coincide in general.
The proof of \cite[Theorem~1]{YHD2006} demonstrates
achievability of all rates in
the rectangular part of Receiver 1's $(R_1, R_2)$ region given in our Theorem~\ref{thm:sup-coding-inner-bound}.
With our simultaneous decoding non-unique decoding strategy, we
can achieve any rate in the triangular part of this region
as well, which could be useful if
the first constraint above on Receiver~2 is looser
than the second constraint above on Receiver~2.
In such a case, the successive decoding
strategy from \cite[Theorem 1]{YHD2006} would not be able to achieve
the rate $R_2$ if $R_2 > I(W;B_1)$, but the simultaneous decoding strategy can.
It might be the case that the proof of \cite[Theorem~1]{YHD2006} could be completed 
by choosing particular coding distributions and taking unions over the resulting regions,
but this is not discussed there.
\end{remark}

\section{Marton coding scheme}		\label{sec:marton}

	We now prove that the Marton inner bound is achievable
	for quantum broadcast channels.
	The Marton scheme depends on auxiliary random variables $U_1$ and $U_2$, 
	\emph{multicoding}, and the properties of strongly typical sequences and projectors.
	The proof depends on some ideas originally presented in \cite{SW12ISIT} and critically on the
	``overcounting'' technique recently presented in \cite{RSW14}.
	\begin{theorem}[Marton inner bound] \label{thm:marton-no-common}
		Let %
		$\{\rho^{B_1B_2}_x \}$ %
		be a classical-quantum broadcast channel and $x=f(u_1,u_2)$ be a deterministic function.
		The following rate region is achievable:
		\bea
			R_1 	&\leq&	I(U_1; B_1)_\theta,   \\
			R_2	&\leq&	I(U_2; B_2)_\theta,  \\ 
		R_1+R_2	&\leq&	I(U_1; B_1)_\theta + I(U_2; B_2)_\theta - I(U_1; U_2)_\theta,  
		\eea			
		where the information quantities are with respect to the state:
		\begin{equation}
			\theta^{U_1U_2B_1B_2} 
			=
			\sum_{u_1,u_2} 
				p(u_1,u_2) 
				\ketbra{u_1}{u_1}^{U_1}
				\otimes
				\ketbra{u_2}{u_2}^{U_2}
				\otimes
				\rho_{f(u_1,u_2)}^{B_1B_2}.
		\end{equation}
		It suffices to take the cardinalities $\mathcal{U}_1$ and $\mathcal{U}_2$ of $U_1$ and $U_2$ to be no larger than the cardinality of the channel's input alphabet
		$\mathcal{X}$: i.e., $|\mathcal{U}_1|,|\mathcal{U}_2| \leq |\mathcal{X}|$.
	\end{theorem}

Define the following states:%
\begin{align}
\rho_{f\left(  u_{1},u_{2}\right)  }^{B_{1}}  &  \equiv\text{Tr}_{B_{2}%
}\left\{  \rho_{f\left(  u_{1},u_{2}\right)  }^{B_{1}B_{2}}\right\}  ,\\
\omega_{u_{1}}^{B_{1}}  &  \equiv\sum_{u_{2}}p_{U_{2}|U_{1}}\!\left(
u_{2}|u_{1}\right)  \rho_{f\left(  u_{1},u_{2}\right)  }^{B_{1}},\\
\overline{\rho}^{B_{1}}  &  \equiv\sum_{u_{1}}p_{U_{1}}\!\left(  u_{1}\right)
\omega_{u_{1}}^{B_{1}}.
\end{align}

\bigskip

\textbf{Codebook construction.} Define two auxiliary indices $l_{1}\in\left\{
1,\ldots,L_{1}\right\}  $ and $l_{2}\in\left\{  1,\ldots,L_{2}\right\}  $, and
let $\widetilde{R}_{1}=\left(  \log L_{1}\right)  /n$ and $\widetilde{R}%
_{2}=\left(  \log L_{2}\right)  /n$. For each $l_{1}$, generate a sequence
$u_{1}^{n}\!\left(  l_{1}\right)  $\ independently and randomly according to the
product distribution%
\begin{equation}
p_{U_{1}^{n}}\!\left(  u_{1}^{n}\right)  \equiv\prod\limits_{i=1}^{n}p_{U_{1}%
}\!\left(  u_{1,i}\right)  .
\end{equation}
Similarly, for each $l_{2}$, generate a sequence $u_{2}^{n}\!\left(
l_{2}\right)  $\ independently and randomly according to the product
distribution%
\begin{equation}
p_{U_{2}^{n}}\!\left(  u_{2}^{n}\right)  \equiv\prod\limits_{i=1}^{n}p_{U_{2}}\!\left(  u_{2,i}\right)  .
\end{equation}
Partition the sequences $u_{1}^{n}\!\left(  l_{1}\right)  $ into $2^{nR_{1}}$
different bins, each of which we label as $B_{m_{1}}$. Partition the sequences
$u_{2}^{n}\!\left(  l_{2}\right)  $ into $2^{nR_{2}}$ different bins, each of
which we label as $C_{m_{2}}$. For each message pair, the sender selects a
sequence pair $\left(  u_{1}^{n}\!\left(  l_{1}\right)  ,u_{2}^{n}\!\left(
l_{2}\right)  \right)  \in\left(  B_{m_{1}}\times C_{m_{2}}\right)
\cap\mathcal{A}_{p_{U_{1},U_{2}},\delta}^{n}$, where $\mathcal{A}%
_{p_{U_{1},U_{2}},\delta}^{n}$ is the strongly typical set for $p_{U_{1}%
,U_{2}}$. The scheme is such that each sequence is taken from the appropriate
bin and the sender demands that they are strongly jointly-typical (otherwise
admitting failure by just sending the first sequence pair in the bin). The
codebook $x^{n}\!\left(  m_{1},m_{2}\right)  $ is deterministically constructed
from $\left(  u_{1}^{n}\!\left(  l_{1}\right)  ,u_{2}^{n}\!\left(  l_{2}\right)
\right)  $, by applying the function $x_{i}=f\!\left(  u_{1,i},u_{2,i}\right)  $.

\bigskip

\textbf{Transmission.} Let $\ell_{1}$ and $\ell_{2}$ denote the pair of
indices of the joint sequence $\left(  u_{1}^{n}\!\left(  \ell_{1}\right)
,u_{2}^{n}\!(  \ell_{2})  \right)  $\ which are chosen as the
codewords for the message pair $\left(  m_{1},m_{2}\right)  $. Expressed in
terms of these indices, the output of the channel is%
\begin{equation}
\rho_{\ell_{1},\ell_{2}}\equiv\bigotimes\limits_{i=1}^{n}\rho_{f\left(
u_{1,i}\left(  \ell_{1}\right)  ,u_{2,i}(  \ell_{2})  \right)
}^{B_{1,i}B_{2,i}}.
\end{equation}
We will also make the abbreviation%
\begin{equation}
\rho_{u_{1}^{n}\left(  \ell_{1}\right)  ,u_{2}^{n}(  \ell_{2})
}^{B_{1}^{n}B_{2}^{n}}\equiv\rho_{\ell_{1},\ell_{2}},
\end{equation}
and furthermore define $\rho_{u_{1}^{n}\left(  \ell_{1}\right)  ,u_{2}%
^{n}(  \ell_{2})  }^{B_{1}^{n}}$ in the obvious way by taking the
partial trace over $B_{2}^{n}$.

\bigskip

\textbf{Decoding.} The decoding POVM\ $\left\{  \Lambda_{l_{1}}\right\}
_{l_{1}\in\left\{  1,\ldots,L_{1}\right\}  }$\ for Receiver 1 is a square-root
measurement as in \eqref{eq:square-root-POVM-generic} and
%\begin{equation}
%\Lambda_{l_{1}}^{B_{1}^{n}}\equiv\left(  \sum_{l_{1}^{\prime}}\Gamma
%_{l_{1}^{\prime}}\right)  ^{-1/2}\Gamma_{l_{1}}\left(  \sum_{l_{1}^{\prime}%
%}\Gamma_{l_{1}^{\prime}}\right)  ^{-1/2}%
%\end{equation}
based on the following positive operators:%
\begin{equation}
\Gamma_{l_{1}}\equiv\Pi_{\overline{\rho},\delta^{\prime\prime}}^{n}\Pi
_{\omega_{u_{1}^{n}\left(  l_{1}\right)  },\delta^{\prime}}\Pi_{\overline
{\rho},\delta^{\prime\prime}}^{n}, \label{eq:sqrt-meas-marton}
\end{equation}
where $\Pi_{\overline{\rho},\delta^{\prime\prime}}^{n}$ is a strongly typical
projector for the state $\overline{\rho}^{\otimes n}$ and $\Pi_{\omega
_{u_{1}^{n}\left(  l_{1}\right)  },\delta^{\prime}}$ is a strong conditionally
typical projector for the state $\omega_{u_{1}^{n}\left(  l_{1}\right)  }$
(cf.~\cite[Chapter 14]{wilde2011book}). Having decoded $l_{1}$ correctly and knowing
the binning scheme, Receiver 1 can deduce the message $m_{1}$ from the bin
index. The decoding is essentially the same for Receiver 2 but using the
appropriate states and induced conditionally typical projectors. Let
$\Lambda_{l_{2}}^{B_{2}^{n}}$ denote the resulting decoding POVM\ for Receiver 2.

\bigskip

\textbf{Error Analysis.} We begin by analyzing the case when $\left(
m_{1},m_{2}\right)  =\left(  1,1\right)  $ and a fixed subcodebook. Let
$\ell_{1}$ and $\ell_{2}$ denote the pair of indices of the joint sequence
$\left(  u_{1}^{n}\left(  \ell_{1}\right)  ,u_{2}^{n}(  \ell_{2})
\right)  $\ which was chosen as the codeword for the message pair $\left(
1,1\right)  $. If there is none, let $\ell_{1}$ and $\ell_{2}$ be the first
pair in the bin. An error occurs if one or more of the following occurs:

\begin{enumerate}
\item Let $\mathcal{E}_{0}$ be the event that $\left(  \left(  u_{1}%
^{n}\!\left(  l_{1}\right)  ,u_{2}^{n}\!\left(  l_{2}\right)  \right)
\notin\mathcal{A}_{p_{U_{1},U_{2}},\delta}^{n}\right)  $ for all $\left(
u_{1}^{n}\!\left(  l_{1}\right)  ,u_{2}^{n}\!\left(  l_{2}\right)  \right)  \in
B_{m_{1}}\times C_{m_{2}}$.

\item The event $\mathcal{E}_{0}^{c}$ occurs, and Receiver 1\ decodes some
other $l_{1}^{\prime}$ besides the transmitted $\ell_{1}$ or Receiver
2\ decodes some other $l_{2}^{\prime}$ besides the transmitted $\ell_{2}$.
\end{enumerate}

We can write out an exact expression for the error probability of a fixed
subcodebook explicitly as%
\begin{equation}
\mathcal{I}\left(  \mathcal{E}_{0}\right)  +\mathcal{I}\left(  \mathcal{E}%
_{0}^{c}\right)  \text{Tr}\left\{  \left(  I^{B_{1}^{n}B_{2}^{n}}%
-\Lambda_{\ell_{1}}^{B_{1}^{n}}\otimes\Lambda_{\ell_{2}}^{B_{2}^{n}}\right)
\rho_{u_{1}^{n}\left(  \ell_{1}\right)  ,u_{2}^{n}(  \ell_{2})
}^{B_{1}^{n}B_{2}^{n}}\right\}  . \label{eq:exact-error}%
\end{equation}
The interpretation is that:

\begin{enumerate}
\item If there are no jointly typical pairs in the subcodebook, then an error
occurs with probability one. So $\mathcal{I}\left(  \mathcal{E}_{0}\right)  $
gives this contribution.

\item If there is at least one pair that is jointly typical (event
$\mathcal{E}_{0}^{c}$ occurs), let $\left(  \ell_{1},\ell_{2}\right)  $ denote
the first one found when scanning in lexiographic order. This one is sent. The
expression for the decoding error probability is exactly equal to%
\begin{equation}
\text{Tr}\left\{  \left(  I^{B_{1}^{n}B_{2}^{n}}-\Lambda_{\ell_{1}}^{B_{1}%
^{n}}\otimes\Lambda_{\ell_{2}}^{B_{2}^{n}}\right)  \rho_{u_{1}^{n}\left(
\ell_{1}\right)  ,u_{2}^{n}(  \ell_{2})  }^{B_{1}^{n}B_{2}^{n}%
}\right\}  .
\end{equation}

\end{enumerate}

\noindent
Note from the \textquotedblleft union bound\textquotedblright\ stated in \eqref{eq:commuting-union-bnd},
we get that%
\begin{multline}
\text{Tr}\left\{  \left(  I^{B_{1}^{n}B_{2}^{n}}-\Lambda_{\ell_{1}}^{B_{1}%
^{n}}\otimes\Lambda_{\ell_{2}}^{B_{2}^{n}}\right)  \rho_{u_{1}^{n}\left(
\ell_{1}\right)  ,u_{2}^{n}(  \ell_{2})  }^{B_{1}^{n} B_2^n}\right\}
\leq\text{Tr}\left\{  \left(  I^{B_{1}^{n}}-\Lambda_{\ell_{1}}^{B_{1}^{n}%
}\right)  \rho_{u_{1}^{n}\left(  \ell_{1}\right)  ,u_{2}^{n}\left(  \ell
_{2}\right)  }^{B_{1}^{n}}\right\}  \\
+\text{Tr}\left\{  \left(  I^{B_{2}^{n}}-\Lambda_{\ell_{2}}^{B_{2}^{n}%
}\right)  \rho_{u_{1}^{n}\left(  \ell_{1}\right)  ,u_{2}^{n}\left(  \ell
_{2}\right)  }^{B_{2}^{n}}\right\}  .
\end{multline}
So this means that we can bound (\ref{eq:exact-error}) from above by%
\begin{equation}
\mathcal{I}\left(  \mathcal{E}_{0}\right)  +\mathcal{I}\left(  \mathcal{E}%
_{0}^{c}\right)  \left[  \text{Tr}\left\{  \left(  I^{B_{1}^{n}}-\Lambda
_{\ell_{1}}^{B_{1}^{n}}\right)  \rho_{u_{1}^{n}\left(  \ell_{1}\right)
,u_{2}^{n}(  \ell_{2})  }^{B_{1}^{n}}\right\}  +\text{Tr}\left\{
\left(  I^{B_{2}^{n}}-\Lambda_{\ell_{2}}^{B_{2}^{n}}\right)  \rho_{u_{1}%
^{n}\left(  \ell_{1}\right)  ,u_{2}^{n}(  \ell_{2})  }^{B_{2}^{n}%
}\right\}  \right]  \label{eq:exact-error-comm-union}%
\end{equation}
We now focus on the term%
\begin{equation}
\text{Tr}\left\{  \left(  I^{B_{1}^{n}}-\Lambda_{\ell_{1}}^{B_{1}^{n}}\right)
\rho_{u_{1}^{n}\left(  \ell_{1}\right)  ,u_{2}^{n}(  \ell_{2})
}^{B_{1}^{n}}\right\}  .\label{eq:exact-error-rec-1}%
\end{equation}
By applying the Hayashi-Nagaoka operator inequality, we can bound
(\ref{eq:exact-error-rec-1}) from above by%
\begin{multline}
2\text{Tr}\left\{  \left(  I-\Pi_{\overline{\rho},\delta^{\prime\prime}}%
^{n}\Pi_{\omega_{u_{1}^{n}\left(  \ell_{1}\right)  },\delta^{\prime}}%
\Pi_{\overline{\rho},\delta^{\prime\prime}}^{n}\right)  \rho_{u_{1}^{n}\left(
\ell_{1}\right)  ,u_{2}^{n}(  \ell_{2})  }^{B_{1}^{n}}\right\}
\label{eq:exact-error-unraveling}\\
+4\sum_{l_{1}^{\prime}\in B_{m_{1}}\wedge l_{1}^{\prime}\neq\ell_{1}}%
\text{Tr}\left\{  \Pi_{\overline{\rho},\delta^{\prime\prime}}^{n}\Pi
_{\omega_{u_{1}^{n}(l_{1}^{\prime})  },\delta^{\prime}}%
\Pi_{\overline{\rho},\delta^{\prime\prime}}^{n}\rho_{u_{1}^{n}\left(  \ell
_{1}\right)  ,u_{2}^{n}(  \ell_{2})  }^{B_{1}^{n}}\right\}  \\
4\sum_{l_{1}^{\prime}\notin B_{m_{1}}}\text{Tr}\left\{  \Pi_{\overline{\rho
},\delta^{\prime\prime}}^{n}\Pi_{\omega_{u_{1}^{n}(  l_{1}^{\prime})  },\delta^{\prime}}\Pi_{\overline{\rho},\delta^{\prime\prime}}%
^{n}\rho_{u_{1}^{n}\left(  \ell_{1}\right)  ,u_{2}^{n}(  \ell_{2})
}^{B_{1}^{n}}\right\}  .
\end{multline}
Consider that%
\begin{align}
&\pullleft\pullleft\text{Tr}\left\{  \Pi_{\overline{\rho},\delta^{\prime\prime}}^{n}%
\Pi_{\omega_{u_{1}^{n}\left(  \ell_{1}\right)  },\delta^{\prime}}%
\Pi_{\overline{\rho},\delta^{\prime\prime}}^{n}\rho_{u_{1}^{n}\left(  \ell
_{1}\right)  ,u_{2}^{n}(  \ell_{2})  }^{B_{1}^{n}}\right\}
\nonumber\\
&  \geq\text{Tr}\left\{  \Pi_{\omega_{u_{1}^{n}\left(  \ell_{1}\right)
},\delta^{\prime}}\rho_{u_{1}^{n}\left(  \ell_{1}\right)  ,u_{2}^{n}\left(
\ell_{2}\right)  }^{B_{1}^{n}}\right\}  -\left\Vert \rho_{u_{1}^{n}\left(
\ell_{1}\right)  ,u_{2}^{n}(  \ell_{2})  }^{B_{1}^{n}}%
-\Pi_{\overline{\rho},\delta^{\prime\prime}}^{n}\rho_{u_{1}^{n}\left(
\ell_{1}\right)  ,u_{2}^{n}(  \ell_{2})  }^{B_{1}^{n}}%
\Pi_{\overline{\rho},\delta^{\prime\prime}}^{n}\right\Vert _{1}\\
&  \geq1-\varepsilon-2\sqrt{\varepsilon^{\prime}}%
\end{align}
where these inequalities follow from Lemma~\ref{lemma-avg-typ-proj-works} in the appendix,
whenever $u_{1}^{n}\!\left(  \ell_{1}\right)  ,u_{2}^{n}\!(  \ell_{2})
$ are strongly jointly typical. We would like to remove the dependence of the
second term in (\ref{eq:exact-error-unraveling}) on the chosen indices
$\ell_{1}$ and $\ell_{2}$. In order to do so, we bound the relevant expression
using the ``overcounting'' idea from \cite{RSW14}:%
\begin{align}
&\pullleft\pullleft\sum_{l_{1}^{\prime}\in B_{m_{1}}\wedge l_{1}^{\prime}\neq\ell_{1}%
}\text{Tr}\left\{  \Pi_{\overline{\rho},\delta^{\prime\prime}}^{n}\Pi
_{\omega_{u_{1}^{n}(l_{1}^{\prime})  },\delta^{\prime}}%
\Pi_{\overline{\rho},\delta^{\prime\prime}}^{n}\rho_{u_{1}^{n}(\ell_{1})  ,u_{2}^{n}(  \ell_{2})  }^{B_{1}^{n}}\right\}  \\
&  =\sum_{l_{1}\in B_{m_{1}},l_{2}\in C_{m_{2}}}\mathcal{I}\left(  l_{1}=\ell_{1},l_{2}=\ell_{2}\right)  \sum_{l_{1}^{\prime}\in B_{m_{1}}\wedge
l_{1}^{\prime}\neq\ell_{1}}\text{Tr}\left\{  \Pi_{\overline{\rho}%
,\delta^{\prime\prime}}^{n}\Pi_{\omega_{u_{1}^{n}(  l_{1}^{\prime})  },\delta^{\prime}}\Pi_{\overline{\rho},\delta^{\prime\prime}}%
^{n}\rho_{u_{1}^{n}\left(  \ell_{1}\right)  ,u_{2}^{n}(  \ell_{2})
}^{B_{1}^{n}}\right\}  \\
&  =\sum_{l_{1}\in B_{m_{1}},l_{2}\in C_{m_{2}}}\mathcal{I}\left(  l_{1}%
=\ell_{1},l_{2}=\ell_{2}\right)  \sum_{l_{1}^{\prime}\in B_{m_{1}}\wedge
l_{1}^{\prime}\neq l_{1}}\text{Tr}\left\{  \Pi_{\overline{\rho},\delta
^{\prime\prime}}^{n}\Pi_{\omega_{u_{1}^{n}(l_{1}^{\prime})
},\delta^{\prime}}\Pi_{\overline{\rho},\delta^{\prime\prime}}^{n}\rho
_{u_{1}^{n}\left(  l_{1}\right)  ,u_{2}^{n}\left(  l_{2}\right)  }^{B_{1}^{n}%
}\right\}  \\
&  \leq\sum_{l_{1}\in B_{m_{1}},l_{2}\in C_{m_{2}}}\mathcal{I}\left(  \left(
l_{1},l_{2}\right)  \in\mathcal{A}\right)  \sum_{l_{1}^{\prime}\in B_{m_{1}%
}\wedge l_{1}^{\prime}\neq l_{1}}\text{Tr}\left\{  \Pi_{\overline{\rho}%
,\delta^{\prime\prime}}^{n}\Pi_{\omega_{u_{1}^{n}(  l_{1}^{\prime})  },\delta^{\prime}}\Pi_{\overline{\rho},\delta^{\prime\prime}}%
^{n}\rho_{u_{1}^{n}\left(  l_{1}\right)  ,u_{2}^{n}\left(  l_{2}\right)
}^{B_{1}^{n}}\right\}  \\
&  =\sum_{l_{1}\in B_{m_{1}},l_{2}\in C_{m_{2}}}\sum_{l_{1}^{\prime}\in
B_{m_{1}}\wedge l_{1}^{\prime}\neq l_{1}}\mathcal{I}\left(  \left(
l_{1},l_{2}\right)  \in\mathcal{A}\right)  \text{Tr}\left\{  \Pi
_{\overline{\rho},\delta^{\prime\prime}}^{n}\Pi_{\omega_{u_{1}^{n}(l_{1}^{\prime})  },\delta^{\prime}}\Pi_{\overline{\rho},\delta
^{\prime\prime}}^{n}\rho_{u_{1}^{n}\left(  l_{1}\right)  ,u_{2}^{n}(l_{2})  }^{B_{1}^{n}}\right\}
\end{align}
where $\mathcal{I}\!\left(  \left(  l_{1},l_{2}\right)  \in\mathcal{A}\right)  $
is a shorthand for $\mathcal{I}\left(  \left(  u_{1}^{n}(l_{1})
,u_{2}^{n}(  l_{2})  \right)  \in\mathcal{A}_{p_{U_{1},U_{2}%
},\delta}^{n}\right)  $. So the final bound on (\ref{eq:exact-error-rec-1}) is%
\begin{multline}
2\left(  \varepsilon+2\sqrt{\varepsilon^{\prime}}\right)  +4\sum_{l_{1}\in
B_{m_{1}},l_{2}\in C_{m_{2}}}\sum_{l_{1}^{\prime}\in B_{m_{1}}\wedge
l_{1}^{\prime}\neq l_{1}}\mathcal{I}\left(  \left(  l_{1},l_{2}\right)
\in\mathcal{A}\right)  \text{Tr}\left\{  \Pi_{\overline{\rho},\delta
^{\prime\prime}}^{n}\Pi_{\omega_{u_{1}^{n}(l_{1}^{\prime})
},\delta^{\prime}}\Pi_{\overline{\rho},\delta^{\prime\prime}}^{n}\rho
_{u_{1}^{n}\left(  l_{1}\right)  ,u_{2}^{n}\left(  l_{2}\right)  }^{B_{1}^{n}%
}\right\}  \\
+4\sum_{l_{1}^{\prime}\notin B_{m_{1}}}\text{Tr}\left\{  \Pi_{\overline{\rho
},\delta^{\prime\prime}}^{n}\Pi_{\omega_{u_{1}^{n}(  l_{1}^{\prime})  },\delta^{\prime}}\Pi_{\overline{\rho},\delta^{\prime\prime}}%
^{n}\rho_{u_{1}^{n}\left(  \ell_{1}\right)  ,u_{2}^{n}(  \ell_{2})
}^{B_{1}^{n}}\right\}  .
\end{multline}
In a similar way, we can write a bound on the right hand term in
(\ref{eq:exact-error-comm-union}) as follows:%
\begin{multline}
\text{Tr}\left\{  \left(  I^{B_{2}^{n}}-\Lambda_{\ell_{2}}^{B_{2}^{n}}\right)
\rho_{u_{1}^{n}\left(  \ell_{1}\right)  ,u_{2}^{n}(  \ell_{2})
}^{B_{2}^{n}}\right\}  \leq2\left(  \varepsilon+2\sqrt{\varepsilon^{\prime}%
}\right)  \\
+4\sum_{l_{1}\in B_{m_{1}},l_{2}\in C_{m_{2}}}\sum_{l_{2}^{\prime}\in
C_{m_{2}}\wedge l_{2}^{\prime}\neq l_{2}}\mathcal{I}\left(  \left(
l_{1},l_{2}\right)  \in\mathcal{A}\right)  \text{Tr}\left\{  \Pi
_{\overline{\overline{\rho}},\delta^{\prime\prime}}^{n}\Pi_{\omega_{u_{2}%
^{n}(  l_{2}^{\prime})  },\delta^{\prime}}\Pi_{\overline
{\overline{\rho}},\delta^{\prime\prime}}^{n}\rho_{u_{1}^{n}\left(
l_{1}\right)  ,u_{2}^{n}\left(  l_{2}\right)  }^{B_{2}^{n}}\right\}  \\
+4\sum_{l_{2}^{\prime}\notin C_{m_{2}}}\text{Tr}\left\{  \Pi_{\overline
{\overline{\rho}},\delta^{\prime\prime}}^{n}\Pi_{\omega_{u_{2}^{n}(l_{2}^{\prime})  },\delta^{\prime}}\Pi_{\overline{\overline{\rho}%
},\delta^{\prime\prime}}^{n}\rho_{u_{1}^{n}\left(  \ell_{1}\right)  ,u_{2}%
^{n}(  \ell_{2})  }^{B_{2}^{n}}\right\}  ,
\end{multline}
where the typical projectors $\Pi_{\overline{\overline{\rho}},\delta
^{\prime\prime}}^{n}$ and $\Pi_{\omega_{u_{2}^{n}(  l_{2}^{\prime})  },\delta^{\prime}}$ acting on system $B_{2}^{n}$ are defined from
the states%
\begin{align}
\omega_{u_{2}}^{B_{2}} &  \equiv\sum_{u_{1}}p_{U_{1}|U_{2}}\!\left(  u_{1}%
|u_{2}\right)  \rho_{f\left(  u_{1},u_{2}\right)  }^{B_{2}},\\
\overline{\overline{\rho}}^{B_{2}} &  \equiv\sum_{u_{2}}p_{U_{2}}\!\left(
u_{2}\right)  \omega_{u_{2}}^{B_{2}}.
\end{align}

\noindent
Putting everything together, we can write a bound on (\ref{eq:exact-error}) as
follows:%
\begin{multline}
\mathcal{I}\left(  \mathcal{E}_{0}\right)  +\mathcal{I}\left(  \mathcal{E}%
_{0}^{c}\right)  \text{Tr}\left\{  \left(  I^{B_{1}^{n}B_{2}^{n}}%
-\Lambda_{\ell_{1}}^{B_{1}^{n}}\otimes\Lambda_{\ell_{2}}^{B_{2}^{n}}\right)
\rho_{u_{1}^{n}\left(  \ell_{1}\right)  ,u_{2}^{n}(  \ell_{2})
}^{B_{1}^{n}B_{2}^{n}}\right\}  \leq\mathcal{I}\left(  \mathcal{E}_{0}\right)
+4\left(  \varepsilon+2\sqrt{\varepsilon^{\prime}}\right)
\label{eq:subcodebook-bound}\\[2mm]
+4\sum_{l_{1}\in B_{m_{1}},l_{2}\in C_{m_{2}}}\sum_{l_{1}^{\prime}\in
B_{m_{1}}\wedge l_{1}^{\prime}\neq l_{1}}\mathcal{I}\left(  \left(
l_{1},l_{2}\right)  \in\mathcal{A}\right)  \text{Tr}\left\{  \Pi
_{\overline{\rho},\delta^{\prime\prime}}^{n}\Pi_{\omega_{u_{1}^{n}(l_{1}^{\prime})  },\delta^{\prime}}\Pi_{\overline{\rho},\delta
^{\prime\prime}}^{n}\rho_{u_{1}^{n}\left(  l_{1}\right)  ,u_{2}^{n}\left(
l_{2}\right)  }^{B_{1}^{n}}\right\}  \\
+4\sum_{l_{1}^{\prime}\notin B_{m_{1}}}\text{Tr}\left\{  \Pi_{\overline{\rho
},\delta^{\prime\prime}}^{n}\Pi_{\omega_{u_{1}^{n}(  l_{1}^{\prime})  },\delta^{\prime}}\Pi_{\overline{\rho},\delta^{\prime\prime}}%
^{n}\rho_{u_{1}^{n}\left(  \ell_{1}\right)  ,u_{2}^{n}(  \ell_{2})
}^{B_{1}^{n}}\right\}  \\
+4\sum_{l_{1}\in B_{m_{1}},l_{2}\in C_{m_{2}}}\sum_{l_{2}^{\prime}\in
C_{m_{2}}\wedge l_{2}^{\prime}\neq l_{2}}\mathcal{I}\left(  \left(
l_{1},l_{2}\right)  \in\mathcal{A}\right)  \text{Tr}\left\{  \Pi
_{\overline{\overline{\rho}},\delta^{\prime\prime}}^{n}\Pi_{\omega_{u_{2}%
^{n}(  l_{2}^{\prime})  },\delta^{\prime}}\Pi_{\overline
{\overline{\rho}},\delta^{\prime\prime}}^{n}\rho_{u_{1}^{n}\left(
l_{1}\right)  ,u_{2}^{n}\left(  l_{2}\right)  }^{B_{2}^{n}}\right\}  \\
+4\sum_{l_{2}^{\prime}\notin C_{m_{2}}}\text{Tr}\left\{  \Pi_{\overline
{\overline{\rho}},\delta^{\prime\prime}}^{n}\Pi_{\omega_{u_{2}^{n}(l_{2}^{\prime})  },\delta^{\prime}}\Pi_{\overline{\overline{\rho}%
},\delta^{\prime\prime}}^{n}\rho_{u_{1}^{n}\left(  \ell_{1}\right)  ,u_{2}%
^{n}(  \ell_{2})  }^{B_{2}^{n}}\right\}  .
\end{multline}

\noindent
At this point, we follow Shannon and recognize that the analysis of a
particular subcode can be difficult (the last four terms above), so we instead
analyze the expectation of the error probability, where the expectation is
with respect to a randomly chosen code. That is, we consider the following
quantity instead%
\begin{equation}
\mathbb{E}_{\mathcal{C}}\left\{  \mathcal{I}\left(  \mathcal{E}_{0}\right)
+\mathcal{I}\left(  \mathcal{E}_{0}^{c}\right)  \text{Tr}\left\{  \left(
I^{B_{1}^{n}B_{2}^{n}}-\Lambda_{\ell_{1}}^{B_{1}^{n}}\otimes\Lambda_{\ell_{2}%
}^{B_{2}^{n}}\right)  \rho_{U_{1}^{n}\left(  \ell_{1}\right)  ,U_{2}%
^{n}(  \ell_{2})  }^{B_{1}^{n}B_{2}^{n}}\right\}  \right\}
,\label{eq:exact-expected-error}%
\end{equation}
where%
\begin{align}
\mathbb{E}_{\mathcal{C}}\left\{  \cdot\right\}   &  \equiv\mathbb{E}%
_{U_{1}^{n}\left(  1\right)  }\cdots\mathbb{E}_{U_{1}^{n}\left(  L_{1}\right)
}\mathbb{E}_{U_{2}^{n}\left(  1\right)  }\cdots\mathbb{E}_{U_{2}^{n}\left(
L_{2}\right)  }\left\{  \cdot\right\},  \\
\mathbb{E}_{U_{1}^{n}\left(  1\right)  }\left\{  \cdot\right\}   &
=\sum_{u_{1}^{n}\left(  1\right)  }p_{U_{1}^{n}}\left(  u_{1}^{n}\left(
1\right)  \right)  \left\{  \cdot\right\},
\end{align}
with the other expectations defined similarly. Then using the bound from
(\ref{eq:subcodebook-bound}), we find the following bound on
(\ref{eq:exact-expected-error})%
\begin{multline}
\mathbb{E}_{\mathcal{C}}\left\{  \mathcal{I}\left(  \mathcal{E}_{0}\right)
\right\}  +4\left(  \varepsilon+2\sqrt{\varepsilon^{\prime}}\right)  \\
+4\mathbb{E}_{\mathcal{C}}\left\{  \sum_{l_{1}\in B_{m_{1}},l_{2}\in C_{m_{2}%
}}\sum_{l_{1}^{\prime}\in B_{m_{1}}\wedge l_{1}^{\prime}\neq l_{1}}%
\mathcal{I}\left(  \left(  l_{1},l_{2}\right)  \in\mathcal{A}\right)
\text{Tr}\left\{  \Pi_{\overline{\rho},\delta^{\prime\prime}}^{n}\Pi
_{\omega_{U_{1}^{n}(l_{1}^{\prime})  },\delta^{\prime}}%
\Pi_{\overline{\rho},\delta^{\prime\prime}}^{n}\rho_{U_{1}^{n}\left(
l_{1}\right)  ,U_{2}^{n}\left(  l_{2}\right)  }^{B_{1}^{n}}\right\}  \right\}
\\
+4\mathbb{E}_{\mathcal{C}}\left\{  \sum_{l_{1}^{\prime}\notin B_{m_{1}}%
}\text{Tr}\left\{  \Pi_{\overline{\rho},\delta^{\prime\prime}}^{n}\Pi
_{\omega_{U_{1}^{n}(l_{1}^{\prime})  },\delta^{\prime}}%
\Pi_{\overline{\rho},\delta^{\prime\prime}}^{n}\rho_{U_{1}^{n}\left(  \ell
_{1}\right)  ,U_{2}^{n}(  \ell_{2})  }^{B_{1}^{n}}\right\}
\right\}  \\
+4\mathbb{E}_{\mathcal{C}}\left\{  \sum_{l_{1}\in B_{m_{1}},l_{2}\in C_{m_{2}%
}}\sum_{l_{2}^{\prime}\in C_{m_{2}}\wedge l_{2}^{\prime}\neq l_{2}}%
\mathcal{I}\left(  \left(  l_{1},l_{2}\right)  \in\mathcal{A}\right)
\text{Tr}\left\{  \Pi_{\overline{\overline{\rho}},\delta^{\prime\prime}}%
^{n}\Pi_{\omega_{U_{2}^{n}(  l_{2}^{\prime})  },\delta^{\prime}}%
\Pi_{\overline{\overline{\rho}},\delta^{\prime\prime}}^{n}\rho_{U_{1}%
^{n}\left(  l_{1}\right)  ,U_{2}^{n}\left(  l_{2}\right)  }^{B_{2}^{n}%
}\right\}  \right\}  \\
+4\mathbb{E}_{\mathcal{C}}\left\{  \sum_{l_{2}^{\prime}\notin C_{m_{2}}%
}\text{Tr}\left\{  \Pi_{\overline{\overline{\rho}},\delta^{\prime\prime}}%
^{n}\Pi_{\omega_{U_{2}^{n}(  l_{2}^{\prime})  },\delta^{\prime}}%
\Pi_{\overline{\overline{\rho}},\delta^{\prime\prime}}^{n}\rho_{U_{1}%
^{n}\left(  \ell_{1}\right)  ,U_{2}^{n}(  \ell_{2})  }^{B_{2}^{n}%
}\right\}  \right\}  .
\end{multline}
We focus on the second expectation above and can write%
\begin{multline}
\mathbb{E}_{\mathcal{C}}\left\{  \sum_{l_{1}\in B_{m_{1}},l_{2}\in C_{m_{2}}%
}\sum_{l_{1}^{\prime}\in B_{m_{1}}\wedge l_{1}^{\prime}\neq l_{1}}%
\mathcal{I}\left(  \left(  l_{1},l_{2}\right)  \in\mathcal{A}\right)
\text{Tr}\left\{  \Pi_{\overline{\rho},\delta^{\prime\prime}}^{n}\Pi
_{\omega_{U_{1}^{n}(l_{1}^{\prime})  },\delta^{\prime}}%
\Pi_{\overline{\rho},\delta^{\prime\prime}}^{n}\rho_{U_{1}^{n}\left(
l_{1}\right)  ,U_{2}^{n}\left(  l_{2}\right)  }^{B_{1}^{n}}\right\}  \right\}
\label{eq:err-term-same-bin}\\
=\sum_{\substack{l_{1}\in B_{m_{1}},\\l_{2}\in C_{m_{2}}}}\sum
_{\substack{l_{1}^{\prime}\in B_{m_{1}},\\l_{1}^{\prime}\neq l_{1}}%
}\mathbb{E}_{U_{1}^{n}(l_{1}^{\prime})  }\mathbb{E}_{U_{1}%
^{n}\left(  l_{1}\right)  }\mathbb{E}_{U_{2}^{n}\left(  l_{2}\right)
}\text{Tr}\left\{  \Pi_{\overline{\rho},\delta^{\prime\prime}}^{n}\Pi
_{\omega_{U_{1}^{n}(l_{1}^{\prime})  },\delta^{\prime}}%
\Pi_{\overline{\rho},\delta^{\prime\prime}}^{n}\rho_{U_{1}^{n}\left(
l_{1}\right)  ,U_{2}^{n}\left(  l_{2}\right)  }^{B_{1}^{n}}\mathcal{I}\left(
\left(  l_{1},l_{2}\right)  \in\mathcal{A}\right)  \right\}
\end{multline}
We focus on bounding the expression inside the sum, keeping in mind that
$l_{1}^{\prime}\neq l_{1}$:%
\begin{align}
&\pullleft\mathbb{E}_{U_{1}^{n}(l_{1}^{\prime})  }\mathbb{E}_{U_{1}%
^{n}\left(  l_{1}\right)  }\mathbb{E}_{U_{2}^{n}\left(  l_{2}\right)
}\text{Tr}\left\{  \Pi_{\overline{\rho},\delta^{\prime\prime}}^{n}\Pi
_{\omega_{U_{1}^{n}(l_{1}^{\prime})  },\delta^{\prime}}%
\Pi_{\overline{\rho},\delta^{\prime\prime}}^{n}\rho_{U_{1}^{n}\left(
l_{1}\right)  ,U_{2}^{n}\left(  l_{2}\right)  }^{B_{1}^{n}}\mathcal{I}\left(
\left(  l_{1},l_{2}\right)  \in\mathcal{A}\right)  \right\}  \\
&  \leq2^{n\left[  H\left(  B_{1}|U_{1}\right)  _{\theta}+c\delta^{\prime
}\right]  }\mathbb{E}_{U_{1}^{n}(l_{1}^{\prime})  }%
\mathbb{E}_{U_{1}^{n}\left(  l_{1}\right)  }\mathbb{E}_{U_{2}^{n}\left(
l_{2}\right)  }\left\{  \text{Tr}\left\{  \Pi_{\overline{\rho},\delta
^{\prime\prime}}^{n}\omega_{U_{1}^{n}(l_{1}^{\prime})  }%
\Pi_{\overline{\rho},\delta^{\prime\prime}}^{n}\rho_{U_{1}^{n}\left(
l_{1}\right)  ,U_{2}^{n}\left(  l_{2}\right)  }^{B_{1}^{n}}\mathcal{I}\left(
\left(  l_{1},l_{2}\right)  \in\mathcal{A}\right)  \right\}  \right\}  \\
&  =2^{n\left[  H\left(  B_{1}|U_{1}\right)  _{\theta}+c\delta^{\prime
}\right]  }\text{Tr}\left\{  \Pi_{\overline{\rho},\delta^{\prime\prime}}%
^{n}\mathbb{E}_{U_{1}^{n}(l_{1}^{\prime})  }\left\{
\omega_{U_{1}^{n}(l_{1}^{\prime})  }\right\}  \Pi_{\overline
{\rho},\delta^{\prime\prime}}^{n}\mathbb{E}_{U_{1}^{n}\left(  l_{1}\right)
}\mathbb{E}_{U_{2}^{n}\left(  l_{2}\right)  }\left\{  \rho_{U_{1}^{n}\left(
l_{1}\right)  ,U_{2}^{n}\left(  l_{2}\right)  }^{B_{1}^{n}}\mathcal{I}\left(
\left(  l_{1},l_{2}\right)  \in\mathcal{A}\right) \! \right\}  \!\right\}  \\
&  =2^{n\left[  H\left(  B_{1}|U_{1}\right)  _{\theta}+c\delta^{\prime
}\right]  }\text{Tr}\left\{  \Pi_{\overline{\rho},\delta^{\prime\prime}}%
^{n}\overline{\rho}^{\otimes n}\Pi_{\overline{\rho},\delta^{\prime\prime}}%
^{n}\mathbb{E}_{U_{1}^{n}\left(  l_{1}\right)  }\mathbb{E}_{U_{2}^{n}\left(
l_{2}\right)  }\left\{  \rho_{U_{1}^{n}\left(  l_{1}\right)  ,U_{2}^{n}\left(
l_{2}\right)  }^{B_{1}^{n}}\mathcal{I}\left(  \left(  l_{1},l_{2}\right)
\in\mathcal{A}\right)  \right\}  \right\}  \\
&  \leq2^{n\left[  H\left(  B_{1}|U_{1}\right)  _{\theta}+c\delta^{\prime
}\right]  }2^{-n\left[  H\left(  B_{1}\right)  _{\theta}-c^{\prime}%
\delta^{\prime\prime}\right]  }\text{Tr}\left\{  \Pi_{\overline{\rho}%
,\delta^{\prime\prime}}^{n}\mathbb{E}_{U_{1}^{n}\left(  l_{1}\right)
}\mathbb{E}_{U_{2}^{n}\left(  l_{2}\right)  }\left\{  \rho_{U_{1}^{n}\left(
l_{1}\right)  ,U_{2}^{n}\left(  l_{2}\right)  }^{B_{1}^{n}}\mathcal{I}\left(
\left(  l_{1},l_{2}\right)  \in\mathcal{A}\right)  \right\}  \right\}  \\
&  =2^{-n\left[  I\left(  U_{1};B_{1}\right)  _{\theta}-c\delta^{\prime
}-c^{\prime}\delta^{\prime\prime}\right]  }\mathbb{E}_{U_{1}^{n}\left(
l_{1}\right)  }\mathbb{E}_{U_{2}^{n}\left(  l_{2}\right)  }\left\{
\text{Tr}\left\{  \Pi_{\overline{\rho},\delta^{\prime\prime}}^{n}\rho
_{U_{1}^{n}\left(  l_{1}\right)  ,U_{2}^{n}\left(  l_{2}\right)  }^{B_{1}^{n}%
}\mathcal{I}\left(  \left(  l_{1},l_{2}\right)  \in\mathcal{A}\right)
\right\}  \right\}  \\
&  \leq2^{-n\left[  I\left(  U_{1};B_{1}\right)  _{\theta}-c\delta^{\prime
}-c^{\prime}\delta^{\prime\prime}\right]  }\mathbb{E}_{U_{1}^{n}\left(
l_{1}\right)  }\mathbb{E}_{U_{2}^{n}\left(  l_{2}\right)  }\left\{
\mathcal{I}\left(  \left(  l_{1},l_{2}\right)  \in\mathcal{A}\right)
\right\}  \\
&  \leq2^{-n\left[  I\left(  U_{1};B_{1}\right)  _{\theta}-c\delta^{\prime
}-c^{\prime}\delta^{\prime\prime}\right]  }2^{-n\left[  I\left(  U_{1}%
;U_{2}\right)  _{\theta}-c^{\prime\prime}\delta^{\prime\prime\prime}\right]
}\\
&  =2^{-n\left[  I\left(  U_{1};B_{1}\right)  _{\theta}+I\left(  U_{1}%
;U_{2}\right)  _{\theta}-c^{\prime\prime}\delta^{\prime\prime\prime}%
-c^{\prime}\delta^{\prime\prime}-c\delta^{\prime}\right]  }%
\end{align}
We then find that (\ref{eq:err-term-same-bin}) is bounded from above by%
\begin{equation}
\left\vert B_{m_{1}}\right\vert ^{2}\left\vert C_{m_{2}}\right\vert
2^{-n\left[  I\left(  U_{1};B_{1}\right)  _{\theta}+I\left(  U_{1}%
;U_{2}\right)  _{\theta}-c^{\prime\prime}\delta^{\prime\prime\prime}%
-c^{\prime}\delta^{\prime\prime}-c\delta^{\prime}\right]  }\,.%
\end{equation}
A similar analysis for Receiver 2 gives the following bound:%
\begin{multline}
\mathbb{E}_{\mathcal{C}}\left\{  \sum_{l_{1}\in B_{m_{1}},l_{2}\in C_{m_{2}}%
}\sum_{l_{2}^{\prime}\in C_{m_{2}}\wedge l_{2}^{\prime}\neq l_{2}}%
\text{Tr}\left\{  \Pi_{\overline{\overline{\rho}},\delta^{\prime\prime}}%
^{n}\Pi_{\omega_{U_{2}^{n}(  l_{2}^{\prime})  },\delta^{\prime}}%
\Pi_{\overline{\overline{\rho}},\delta^{\prime\prime}}^{n}\rho_{U_{1}%
^{n}\left(  l_{1}\right)  ,U_{2}^{n}\left(  l_{2}\right)  }^{B_{2}^{n}%
}\right\}  \right\}  \\
\leq\left\vert B_{m_{1}}\right\vert \left\vert C_{m_{2}}\right\vert
^{2}2^{-n\left[  I\left(  U_{2};B_{2}\right)  _{\theta}+I\left(  U_{1}%
;U_{2}\right)  _{\theta}-c^{\prime\prime}\delta^{\prime\prime\prime}%
-c^{\prime}\delta^{\prime\prime}-c\delta^{\prime}\right]  }.
\end{multline}
We can again use the same analysis to recover the following bounds (however
observing that the joint random variable $\left(  U_{1}^{n}\left(  \ell
_{1}\right)  ,U_{2}^{n}(  \ell_{2})  \right)  $ is independent of
both $U_{1}^{n}(l_{1}^{\prime})  $ and $U_{2}^{n}(l_{2}^{\prime})$ for $l_{1}^{\prime}\notin B_{m_{1}}$ and
$l_{2}^{\prime}\notin C_{m_{2}}$):
\begin{align}
\mathbb{E}_{\mathcal{C}}\left\{  \sum_{l_{1}^{\prime}\notin B_{m_{1}}%
}\text{Tr}\left\{  \Pi_{\overline{\rho},\delta^{\prime\prime}}^{n}\Pi
_{\omega_{U_{1}^{n}(l_{1}^{\prime})  },\delta^{\prime}}%
\Pi_{\overline{\rho},\delta^{\prime\prime}}^{n}\rho_{U_{1}^{n}\left(  \ell
_{1}\right)  ,U_{2}^{n}(  \ell_{2})  }^{B_{1}^{n}}\right\}
\right\}   &  \leq L_{1}2^{-n\left[  I\left(  U_{1};B_{1}\right)  _{\theta
}-c^{\prime}\delta^{\prime\prime}-c\delta^{\prime}\right]  },\\
\mathbb{E}_{\mathcal{C}}\left\{  \sum_{l_{2}^{\prime}\notin C_{m_{2}}%
}\text{Tr}\left\{  \Pi_{\overline{\overline{\rho}},\delta^{\prime\prime}}%
^{n}\Pi_{\omega_{U_{2}^{n}(  l_{2}^{\prime})  },\delta^{\prime}}%
\Pi_{\overline{\overline{\rho}},\delta^{\prime\prime}}^{n}\rho_{U_{1}%
^{n}\left(  \ell_{1}\right)  ,U_{2}^{n}(  \ell_{2})  }^{B_{2}^{n}%
}\right\}  \right\}   &  \leq L_{2}2^{-n\left[  I\left(  U_{2};B_{2}\right)
_{\theta}-c^{\prime}\delta^{\prime\prime}-c\delta^{\prime}\right]  }.
\end{align}

The term $\mathbb{E}_{\mathcal{C}}\left\{  \mathcal{I}\left(  \mathcal{E}%
_{0}\right)  \right\}  $ can be bounded from above by $\varepsilon$ by
employing the mutual covering lemma \cite{eGvdM81} (see also \cite[Lemma 8.1]{el2010lecture}).
Indeed if we choose%
\begin{equation}
\left(  \widetilde{R}_{1}-R_{1}\right)  +\left(  \widetilde{R}_{2}%
-R_{2}\right)  \geq I\!\left(  U_{1};U_{2}\right)  _{\theta}+\delta
^{\prime\prime\prime\prime},
\end{equation}
then this error can be made arbitrarily small (i.e., less than $\varepsilon$)
by increasing $n$.

So we finally get that (\ref{eq:exact-expected-error}) is bounded from above
by%
\begin{multline}
\varepsilon+4\left(  \varepsilon+2\sqrt{\varepsilon^{\prime}}\right)
+\left\vert B_{m_{1}}\right\vert ^{2}\left\vert C_{m_{2}}\right\vert
2^{-n\left[  I\left(  U_{1};B_{1}\right)  _{\theta}+I\left(  U_{1}%
;U_{2}\right)  _{\theta}-c^{\prime\prime}\delta^{\prime\prime\prime}%
-c^{\prime}\delta^{\prime\prime}-c\delta^{\prime}\right]  }+L_{1}2^{-n\left[
I\left(  U_{1};B_{1}\right)  _{\theta}-c^{\prime}\delta^{\prime\prime}%
-c\delta^{\prime}\right]  }\label{eq:final-err-term}\\
+\left\vert B_{m_{1}}\right\vert \left\vert C_{m_{2}}\right\vert
^{2}2^{-n\left[  I\left(  U_{2};B_{2}\right)  _{\theta}+I\left(  U_{1}%
;U_{2}\right)  _{\theta}-c^{\prime\prime}\delta^{\prime\prime\prime}%
-c^{\prime}\delta^{\prime\prime}-c\delta^{\prime}\right]  }+L_{2}2^{-n\left[
I\left(  U_{2};B_{2}\right)  _{\theta}-c^{\prime}\delta^{\prime\prime}%
-c\delta^{\prime}\right]  }.
\end{multline}
Then for any $\varepsilon^{\prime\prime}>0$, we can pick%
\begin{align}
2\left(  \widetilde{R}_{1}-R_{1}\right)  +\left(  \widetilde{R}_{2}%
-R_{2}\right)   &  <I\!\left(  U_{1};B_{1}\right)  _{\theta}+I\!\left(
U_{1};U_{2}\right)  _{\theta}-c^{\prime\prime}\delta^{\prime\prime\prime
}-c^{\prime}\delta^{\prime\prime}-c\delta^{\prime},\\
\widetilde{R}_{1} &  <I\!\left(  U_{1};B_{1}\right)  _{\theta}-c^{\prime}%
\delta^{\prime\prime}-c\delta^{\prime},\\
\left(  \widetilde{R}_{1}-R_{1}\right)  +2\left(  \widetilde{R}_{2}%
-R_{2}\right)   &  <I\!\left(  U_{2};B_{2}\right)  _{\theta}+I\!\left(
U_{1};U_{2}\right)  _{\theta}-c^{\prime\prime}\delta^{\prime\prime\prime
}-c^{\prime}\delta^{\prime\prime}-c\delta^{\prime},\\
\widetilde{R}_{2} &  <I\!\left(  U_{2};B_{2}\right)  _{\theta}-c^{\prime}%
\delta^{\prime\prime}-c\delta^{\prime},
\end{align}
and $n$ sufficiently large so that the quantity in (\ref{eq:final-err-term})
is less than $\varepsilon^{\prime\prime}$. Indeed this estimate can be made
for the expected error probability for each of the subcodebooks, so by
linearity of the expectation we can finally conclude the existence of a coding
scheme for which%
\begin{equation}
\frac{1}{M_{1}M_{2}}\sum_{m_{1},m_{2}}\text{Tr}\left\{  \left(  I-\Upsilon
_{m_{1}}^{B_{1}^{n}}\otimes\Upsilon_{m_{2}}^{B_{2}^{n}}\right)  \rho
_{m_{1},m_{2}}\right\}  \leq\varepsilon^{\prime\prime}%
\end{equation}
as long as%
\begin{align}
\left(  \widetilde{R}_{1}-R_{1}\right)  +\left(  \widetilde{R}_{2}%
-R_{2}\right)   &  \geq I\!\left(  U_{1};U_{2}\right)  _{\theta}+\delta
^{\prime\prime\prime\prime},\\
2\left(  \widetilde{R}_{1}-R_{1}\right)  +\left(  \widetilde{R}_{2}%
-R_{2}\right)   &  <I\!\left(  U_{1};B_{1}\right)  _{\theta}+I\!\left(
U_{1};U_{2}\right)  _{\theta}-c^{\prime\prime}\delta^{\prime\prime\prime
}-c^{\prime}\delta^{\prime\prime}-c\delta^{\prime},\\
\widetilde{R}_{1} &  <I\!\left(  U_{1};B_{1}\right)  _{\theta}-c^{\prime}%
\delta^{\prime\prime}-c\delta^{\prime},\\
\left(  \widetilde{R}_{1}-R_{1}\right)  +2\left(  \widetilde{R}_{2}%
-R_{2}\right)   &  <I\!\left(  U_{2};B_{2}\right)  _{\theta}+I\!\left(
U_{1};U_{2}\right)  _{\theta}-c^{\prime\prime}\delta^{\prime\prime\prime
}-c^{\prime}\delta^{\prime\prime}-c\delta^{\prime},\\
\widetilde{R}_{2} &  <I\!\left(  U_{2};B_{2}\right)  _{\theta}-c^{\prime}%
\delta^{\prime\prime}-c\delta^{\prime},
\end{align}
where $\Upsilon_{m_{1}}^{B_{1}^{n}}$ and $\Upsilon_{m_{2}}^{B_{2}^{n}}$
represent the overall decoding POVMs of Bob and $\rho_{m_{1},m_{2}}$
represents the channel output when sending messages $m_{1}$ and $m_{2}$. Thus,
since $\delta^{\prime},\delta^{\prime\prime},\delta^{\prime\prime\prime
},\delta^{\prime\prime\prime\prime}>0$ are arbitrary, the following rate
region is achievable:%
\begin{align}
R_{1}+R_{2}+I\!\left(  U_{1};U_{2}\right)  _{\theta} &  \leq\widetilde{R}_{1}+\widetilde{R}_{2}, 									\label{eqf1}	\\
2\widetilde{R}_{1}+\widetilde{R}_{2} &  \leq I\!\left(  U_{1};B_{1}\right)_{\theta}+I\!\left(  U_{1};U_{2}\right)  _{\theta}+2R_{1}+R_{2},		\label{eqf2}	\\
\widetilde{R}_{1} &  \leq I\!\left(  U_{1};B_{1}\right)  _{\theta},															\label{eqf3}	 \\
2\widetilde{R}_{2}+\widetilde{R}_{1} &  \leq I\!\left(  U_{2};B_{2}\right)_{\theta}+I\!\left(  U_{1};U_{2}\right)  _{\theta}+2R_{2}+R_{1},		\label{eqf4}	\\
\widetilde{R}_{2} &  \leq I\!\left(  U_{2};B_{2}\right)  _{\theta}.															\label{eqf5}
\end{align}
By exploiting the additional constraints $R_{1}\leq\widetilde{R}_{1}$ and
$R_{2}\leq\widetilde{R}_{2}$ and applying Fourier-Motzkin elimination (see Appendix~A),
we find that
the following quantum Marton rate region is achievable:%
\begin{align}
R_{1} &  \leq I\!\left(  U_{1};B_{1}\right)  _{\theta},\\
R_{2} &  \leq I\!\left(  U_{2};B_{2}\right)  _{\theta},\\
R_{1}+R_{2} &  \leq I\!\left(  U_{1};B_{1}\right)  _{\theta}+I\!\left(
U_{2};B_{2}\right)  _{\theta}-I\!\left(  U_{1};U_{2}\right)  _{\theta}.
\end{align}

As we are dealing with a channel having a classical input, the cardinality bounds given in the statement in the theorem follow directly from what is known in the classical case. Here, one can apply the perturbation method introduced in \cite{GA12}, discussed also in \cite{JN10}, and reviewed in \cite{el2010lecture}.

\section{Conclusion}	 	\label{sec:conclusion}

	We have proved quantum generalizations of the superposition coding inner bound \cite{C72,B73}
	and the Marton rate region with no common message \cite{M79}.
	A key ingredient in both proofs was the use of 
	the projector trick.
	A natural followup question would be to combine the two strategies to obtain
	the Marton coding scheme with a common message.

	A much broader goal would be to extend all of network information theory to the study of quantum channels. 
	To accomplish this goal, it would be helpful to have a tool that generalizes El Gamal and Kim's classical
	packing lemma \cite{el2010lecture} to the quantum domain. 
	The packing lemma is sufficient to prove all of the known coding theorems in network information theory. 
	At the moment, it is not clear to us whether such a tool exists for the quantum case, 
	but evidence in favor of its existence is that 
	\ 1) one can prove the HSW coding theorem by using conditionally 
	     typical projectors only \cite[Exercise 19.3.5]{wilde2011book}, 
	\ 2) we have solved the quantum simultaneous decoding conjecture for the 
	      case of two senders  \cite{FHSSW11,S11a}, and 
	\ 3) we have generalized two important coding theorems in the current paper 
   	     (with proofs somewhat similar to the classical proofs). 
	Ideally, such a ``quantum packing lemma'' would allow quantum information theorists 
	to prove quantum network coding theorems by appealing to it, 
	rather than having to analyze each coding scheme in detail on a case by case basis.

We acknowledge discussions with Patrick Hayden, Omar Fawzi, Pranab Sen, 
and Saikat Guha during the development of \cite{FHSSW11,S11a}.
We are especially grateful to Andreas Winter for relaying to us the observation
of Pranab Sen that our former argument
for the Marton region was incomplete and to Pranab Sen for subsequent
discussions regarding this observation. We are also very grateful to Jaikumar Radhakrishnan for many
helpful discussions regarding Marton coding and for helping us to understand the overcounting
technique from \cite{RSW14}.
I.~Savov acknowledges support from FQRNT and NSERC.
M.~M.~Wilde acknowledges support from the
Centre de Recherches Math\'ematiques.

\appendices

\section{Fourier-Motzkin elimination}

In the proof of Theorem~\ref{thm:marton-no-common}, we conclude that the following rate region is achievable:%
\begin{align}
R_{1}+R_{2}+I\!\left(  U_{1};U_{2}\right)  _{\theta} &  \leq\widetilde{R}_{1}+\widetilde{R}_{2},				\tag{\ref{eqf1}}	\\
2\widetilde{R}_{1}+\widetilde{R}_{2} &  \leq I\!\left(  U_{1};B_{1}\right)_{\theta}+2R_{1}+R_{2} + I(U_1;U_2) ,	\tag{\ref{eqf2}}	\\
\widetilde{R}_{1} &  \leq I\!\left(  U_{1};B_{1}\right)  _{\theta},										\tag{\ref{eqf3}}	\\
2\widetilde{R}_{2}+\widetilde{R}_{1} &  \leq I\!\left(  U_{2};B_{2}\right)_{\theta}+2R_{2}+R_{1}  + I(U_1;U_2),	\tag{\ref{eqf4}}	\\
\widetilde{R}_{2} &  \leq I\!\left(  U_{2};B_{2}\right)  _{\theta}.										\tag{\ref{eqf5}}
\end{align}
There are additional constraints $R_{1}\leq\widetilde{R}_{1}$ and
$R_{2}\leq\widetilde{R}_{2}$.
To eliminate $\widetilde{R}_{1}$, we split this system of seven equations into three groups:
those that provide lower bounds on $\widetilde{R}_{1}$, those that provide upper bounds on $\widetilde{R}_{1}$, and equations that do not involve $\widetilde{R}_{1}$:
\begin{subequations}
\begin{align}
I(U_1;U_2) + R_{1} + R_{2} - \widetilde{R}_{2} &\leq \widetilde{R}_{1}  \\
R_{1} &\leq \widetilde{R}_{1}  \\
\widetilde{R}_{1} &\leq \frac{I(U_1;B_1)}{2} + R_{1} + \frac{R_{2}}{2} - \frac{\widetilde{R}_{2}}{2}  + \frac{I(U_1;U_2)}{2}\\
\widetilde{R}_{1} &\leq I(U_1;B_1)  \\
\widetilde{R}_{1} &\leq I(U_2;B_2) + R_{1} + 2 R_{2} - 2 \widetilde{R}_{2}  + I(U_1;U_2)\\
\widetilde{R}_{2} &\leq I(U_2;B_2)  \\
R_{2} &\leq \widetilde{R}_{2}  
\end{align}
\end{subequations}
Now we combine each possible lower bound (a,b) with each possible upper bound (c,d,e) and copy over the others:
\begin{subequations}
\begin{align}
I(U_1;U_2) + R_{1} + R_{2} - \widetilde{R}_{2} &\leq \frac{I(U_1;B_1)}{2} + R_{1} + \frac{R_{2}}{2} - \frac{\widetilde{R}_{2}}{2}  +\frac{I(U_1;U_2)}{2}\\
I(U_1;U_2) + R_{1} + R_{2} - \widetilde{R}_{2} &\leq I(U_1;B_1)  \\
I(U_1;U_2) + R_{1} + R_{2} - \widetilde{R}_{2} &\leq I(U_2;B_2) + R_{1} + 2 R_{2} - 2 \widetilde{R}_{2} + I(U_1;U_2) \\
R_{1} &\leq \frac{I(U_1;B_1)}{2} + R_{1} + \frac{R_{2}}{2} - \frac{\widetilde{R}_{2}}{2} +\frac{I(U_1;U_2)}{2}  \\
R_{1} &\leq I(U_1;B_1)  \\
R_{1} &\leq I(U_2;B_2) + R_{1} + 2 R_{2} - 2 \widetilde{R}_{2}  + I(U_1;U_2) \\
\widetilde{R}_{2} &\leq I(U_2;B_2)  \\
R_{2} &\leq \widetilde{R}_{2}  
\end{align}
\end{subequations}
Cancelling terms and simplifying, we get
\begin{subequations}
\label{T1done}
\begin{align}
\frac{I(U_1;U_2)}{2} + \frac{R_{2}}{2} - \frac{\widetilde{R}_{2}}{2} &\leq \frac{I(U_1;B_1)}{2}  \label{eq:redun-1}\\
I(U_1;U_2) + R_{1} + R_{2} - \widetilde{R}_{2} &\leq I(U_1;B_1)  \label{eq:pair-1}\\
0 &\leq I(U_2;B_2) + R_{2} - \widetilde{R}_{2}  \label{eq:redun-3}\\
0 &\leq \frac{I(U_1;B_1)}{2} + \frac{R_{2}}{2} - \frac{\widetilde{R}_{2}}{2}  +\frac{I(U_1;U_2)}{2} \label{eq:redun-2} \\
R_{1} &\leq I(U_1;B_1)  \\
0 &\leq I(U_2;B_2) + 2 R_{2} - 2 \widetilde{R}_{2}  + I(U_1;U_2) \\
\widetilde{R}_{2} &\leq I(U_2;B_2)  \label{eq:pair-2}\\
R_{2} &\leq \widetilde{R}_{2}  
\end{align}
\end{subequations}
This completes the steps required to eliminate $\widetilde{R}_{1}$.

Observe that \eqref{eq:redun-1} is redundant because it is implied by \eqref{eq:pair-1} and
the implicit constraint $R_1 \geq 0$, and \eqref{eq:redun-3} is redundant because it is
implied by \eqref{eq:pair-2} and the
implicit constraint $R_2 \geq 0$. After removing the redundant inequalities,
to eliminate $\widetilde{R}_{2}$, we
rearrange the equations \eqref{T1done} into lower bounds, upper bounds, and those not containing $\widetilde{R}_{2}$:
\begin{subequations}
\begin{align}
- I(U_1;B_1) + I(U_1;U_2) + R_{1} + R_{2} &\leq \widetilde{R}_{2}  \\
R_{2} &\leq \widetilde{R}_{2}  \\
\widetilde{R}_{2} &\leq I(U_1;B_1) + R_{2}   +I(U_1;U_2)  \\
\widetilde{R}_{2} &\leq \frac{I(U_2;B_2)}{2} + R_{2}  + \frac{I(U_1;U_2)}{2}\\
\widetilde{R}_{2} &\leq I(U_2;B_2)  \\
R_{1} &\leq I(U_1;B_1)
\end{align}
\end{subequations}
Combining each of the lower bounds on $\widetilde{R}_{2}$ with each of the upper bounds results in the following equations:
\begin{subequations}
\begin{align}
- I(U_1;B_1) + I(U_1;U_2) + R_{1} + R_{2} &\leq I(U_1;B_1) + R_{2}   +I(U_1;U_2)\\
- I(U_1;B_1) + I(U_1;U_2) + R_{1} + R_{2} &\leq \frac{I(U_2;B_2)}{2} + R_{2}  + \frac{I(U_1;U_2)}{2} \\
- I(U_1;B_1) + I(U_1;U_2) + R_{1} + R_{2} &\leq I(U_2;B_2)  \\
R_{2}&\leq I(U_1;B_1) + R_{2}   +I(U_1;U_2)\\
R_{2} &\leq \frac{I(U_2;B_2)}{2} + R_{2}  + \frac{I(U_1;U_2)}{2} \\
R_{2} &\leq I(U_2;B_2)  \\
R_{1} &\leq I(U_1;B_1)
\end{align}
\end{subequations}
After simplification, this system of equations becomes
\begin{subequations}
\begin{align}
 R_{1}  &\leq 2 I(U_1;B_1)    \\
 R_{1}  &\leq \frac{I(U_2;B_2)}{2} + I(U_1;B_1)  - \frac{I(U_1;U_2)}{2} \label{concern}\\
 R_{1} + R_{2} &\leq I(U_2;B_2) + I(U_1;B_1) - I(U_1;U_2) \label{marton1} \\
 0&\leq I(U_1;B_1)    +I(U_1;U_2)\\
0 &\leq \frac{I(U_2;B_2)}{2}   + \frac{I(U_1;U_2)}{2} \\
R_{2} &\leq I(U_2;B_2)  \\
R_{1} &\leq I(U_1;B_1) \label{marton3}
\end{align}
\end{subequations}

\noindent
Observe that the first inequality is implied by the last and the fourth and fifth inequalities are trivially true. Consider dividing \eqref{marton1} by two, 
	dividing \eqref{marton3} by two, and adding the result:
	\begin{equation}
	R_{1} + \frac{R_2}{2}  \leq \frac{I(U_2;B_2)}{2}  + I(U_1;B_1) - \frac{I(U_1;U_2)}{2} .
	\end{equation}
	Using the fact that $R_2 \geq 0$, we see that \eqref{concern} is redundant and we are left with the three inequalities that specify the Marton region.

\ifthenelse{\boolean{WITHAPDX} }{

\section{Typicality lemma}

The following lemma is an extension of \cite[Property 14.2.7]{wilde2011book}.

	\begin{lemma}	\label{lemma-avg-typ-proj-works}
	The state $\rhoNulul$ is well supported by both the averaged state projector:
	\be
		\Tr\big[ \PIavg \ \rhoNulul \big] \geq 1 - \epsilon, \ \forall \ell_1, \ell_2,
	\ee
	and the $\omgu$ conditionally typical projector:
	\be
		\Tr\left[ \PIone \ \rhoNulul \right] \geq 1 - \epsilon, \  \forall \ell_2,
	\ee
	when $u_1^n(\ell_1)$ and $u_2^n(\ell_2)$ are strongly jointly typical. (Both of these projectors are defined in the main text just after \eqref{eq:sqrt-meas-marton}.)
	\end{lemma}

	Consider the following sets of all jointly-typical and marginally-typical sequences
	for the probability distribution $p_{U_1U_2}(u_1,u_2)$:
	\begin{align*}
		\mcal{A}^n_{p_{U_1},\delta}  
			&\equiv 
			\left\{ u_1^n \in \mcal{U}_1^n \ : \  \left|  \frac{N(u_a|u_1^n)}{n} -  p_{U_1}(u_a) \right| \leq \delta \right\}, \\
		\mcal{A}^n_{p_{U_2},\delta}  
			&\equiv 
			\left\{ u_2^n \in \mcal{U}_2^n \ : \  \left|  \frac{N(u_a|u_2^n)}{n} -  p_{U_2}(u_a) \right| \leq \delta \right\}, \\
		\mcal{A}^n_{p_{U_1U_2},\delta} 
			&\equiv 
			\left\{ u^n \in \mcal{U}_1^n\times\mcal{U}_2^n : \left|  \frac{N(u_a|u^n)}{n} -  p_{U_1U_2}(u_a) \right| \leq \delta \right\}.
	\end{align*}

	Note that the notion of strong typicality implies that if $u^n=(u_1^n,u^n_2) \in \mcal{A}^n_{p_{U_1U_2},\delta}$,
	then both of its substrings are marginally typical: 
	$u_1^n \in \mcal{A}^n_{p_{U_1},\delta}$ and $u_2^n \in \mcal{A}^n_{p_{U_2},\delta}$.

	\begin{IEEEproof}
		Consider the eigen-decomposition of the average state at Receiver 1:
		\be
			\bar{\rho} = \sum_z p_Z(z) \ketbra{z}{z},
		\ee
		and the associated \emph{pinching} operator:
		\be
			\Delta(\psi) = \sum_z \ketbra{z}{z} \psi \ketbra{z}{z},
		\ee
		which turns any quantum state on the output system of Receiver 1
		into a classical probability distribution.
		
		In particular, when $\Delta$ is applied to the state $\rho_{u_1,u_2}$,
		(the channel output  when codewords $u_1$ and $u_2$ are sent) is given by:
		\begin{align}
			\rho^\prime_{u_1,u_2} 
				& =  \sum_z \ketbra{z}{z} \rho_{u_1,u_2} \ketbra{z}{z} \\
				& = \sum_z p_{Z_p|U_1U_2}(z_{p}|u_1,u_2) \ketbra{z}{z},
		\end{align}
		where $p_{Zp|U_1U_2}(z_{p}|u_1,u_2)$ is a classical probability
		distribution.
		
		The statement of the lemma can be expressed in terms of $n$
		copies of this product distribution:
		\begin{align}
			\Tr\left[ \PIavg \ \rhoNulul \right] 
				&= \Tr\left[ \sum_{z^n \in \mcal{A}_{\bar{\rho}} } \ketbra{z^n}{z^n} \ \rhoNulul \right] \\
				&= \Tr\left[ \sum_{z^n \in \mcal{A}_{\bar{\rho}} } \ketbra{z^n}{z^n} \ketbra{z^n}{z^n}  \  \rhoNulul \right]  \\
				&= \Tr\left[ \sum_{z^n \in \mcal{A}_{\bar{\rho}} } \bra{z^n}  \rhoNulul \ket{ z^n} \ \ketbra{z^n}{z^n}  \right]  \\
				& = \sum_{z^n \in \mcal{A}_{\bar{\rho}} } p_{Z_p^n|U^n_1U^n_2}(z_{p}^n|u^n_1,u^n_2) .
		\end{align}
		
		Thus we see that the value of the trace expression is equivalent 
		to the probability of a conditionally typical sequence $Z_p^n|u^n_1u^n_2$ being
		in the typical set $\mcal{A}_{\bar{\rho}} $: 
		\be
			\Pr \left\{  Z_p^n|u^n_1u^n_2 \in \mcal{A}_{\bar{\rho}}  \right\}.
		\ee
		
		To evaluate the above expression we start from the following 
		facts: (1) For $n$ large enough the state classical distribution
		that corresponds to the pinched state $\rho^\prime_{u_1^n,u_2^n}$ is going to be conditionally typical
		on the input sequence:
		\be
			\Pr \left\{  Z_p^n|u^n_1u^n_2 \in \mcal{A} _{Z_p^n|U_1^nU_2^n, \delta'}  \right\} \geq 1 - \eps',
		\ee

		and (2) the input sequence is jointly typical $(u_1^n,u_2^n) \in \mcal{A}_{U_1^nU_2^n, \delta''}$.
		It then follows that, with high probability the input-output sequence will be $(\delta'+\delta'')$-jointly-typical:
		\be
			Z_p^n \in \mcal{A}_{Z_p^nU_1^nU_2^n, \delta' + \delta''},
		\ee
		which in turn implies that:
		\be
			Z_p^n \in \mcal{A}_{Z_p, |\mcal{U}_1||\mcal{U}_2|(\delta' + \delta'')} = \mcal{A}_{\bar{\rho}, |\mcal{U}_1||\mcal{U}_2|(\delta' + \delta'')}.
		\ee
		By a suitable choice of $n$, $\delta'=\frac{\delta}{2|\mcal{U}_1||\mcal{U}_2|}$,
		and $\delta''=\frac{\delta}{2|\mcal{U}_1||\mcal{U}_2|}$ we have that
		\begin{align}
			\Tr\left[ \PIavg \ \rhoNulul \right] 
			 	& = \Pr \left\{  Z_p^n|u^n_1u^n_2 \in \mcal{A}_{\bar{\rho}}  \right\}  \\
				& \geq 1 - \eps''.
		\end{align}

		To prove the other inequality, consider the eigen-decompositions of $\omgu$ states at Receiver 1:
		\be
			\omega_{u_a} = \sum_{z_1} p_{Z_1|U_1}(z_1|u_a) \ketbra{z_1^{(u_a)}}{z_1^{(u_a)}}, \qquad \forall u_a \in \mcal{U}_1.
		\ee
		The associated \emph{pinching} operator is:
		\be
			\Delta_{u_a}(\psi) = \sum_{z_1} \ketbra{z_1^{(u_a)}}{z_1^{(u_a)}} \psi \ketbra{z_1^{(u_a)}}{z_1^{(u_a)}},
		\ee
		which turns any quantum state on the output system of Receiver 1
		into a classical probability distribution expressed in terms of the basis for $\omega_{u_a}$:
		$\ket{z_1^{(u_a)}}$.
		When the symbol $u_a$ is obvious from the context, we will sometimes
		refer to the basis elements simply as $\ket{z_1} \equiv \ket{z_1^{(u_a)}}$.
		
		When $\Delta_{u_1}$ is applied to the state $\rho_{u_1,u_2}$,
		(the channel output  when codewords $u_1$ and $u_2$ are sent) is given by:
		\begin{align}
			\rho^\prime_{u_a,u_b} &= \Delta_{u_a}\left(  \rho_{u_a,u_b} \right) \\
				& =  \sum_{z_1} \ketbra{z_1^{(u_a)}}{z_1^{(u_a)}} \rho_{u_a,u_b} \ketbra{z_1^{(u_a)}}{z_1^{(u_a)}} \\
				& = \sum_{z_1}  p_{Z_{1p}|U_1U_2}(z_{1}|u_a,u_b) \ketbra{z_1^{(u_a)}}{z_1^{(u_a)}},
		\end{align}
		where $p_{Z_{1p}|U_1U_2}(z_{1}|u_a,u_b)$ is a classical probability distribution.

		If we take the conditional marginal of this distribution we get
		\be
			p_{Z_{1p}|U_1}(z_{1}|u_a) 
				 = \sum_{u_b} p_{U_2|U_1}(u_b|u_a) p_{Z_{1p}|U_1U_2}(z_{1}|u_a,u_b),
		\ee
		which is the probability distribution of the $u_a$-basis eigenvalues given that $U_2$ is unknown.
		This distribution can also be obtained from the pinching of the state $\omgu$:
		\begin{align}
			\Delta_{u_a}(\omgu)  \nonumber 
			& =  \sum_{z_1} \ketbra{z_1^{(u_a)}}{z_1^{(u_a)}} \omgu \ketbra{z_1^{(u_a)}}{z_1^{(u_a)}}  \\
			& =  \sum_{z_1} \ketbra{z_1^{(u_a)}}{z_1^{(u_a)}} \sum_{u_b} p_{U_2|U_1}(u_b|u_a) \rho_{u_a,u_b}  \ketbra{z_1^{(u_a)}} {z_1^{(u_a)}}  \\
			& =  \sum_{z_1} \sum_{u_b} p_{U_2|U_1}(u_b|u_a) \underbrace{\bra{z_1^{(u_a)}} \rho_{u_a,u_b} \ket{z_1^{(u_a)}}}_{p_{Z_{1p}|U_1U_2}(z_{1}|u_a,u_b)}  \ketbra{z_1^{(u_a)}} {z_1^{(u_a)}}  \\
			& = \sum_{z_1} p_{Z_{1p}|U_1}(z_{1}|u_a)   \ketbra{z_1^{(u_a)}} {z_1^{(u_a)}} 
		\end{align}
				
		Define the classical conditionally typical sets on sequences of  $m$ symbols  drawn 
		from the above probability distributions:
		\begin{align}
			p_{Z_{1p}|U_1U_2}(z_{1}|u_a,u_b) 
				& \Rightarrow \mcal{A}^{(m)}_{Z_{1p}|u_au_2^m,\delta} \\
			p_{Z_{1p}|U_1}(z_{1}|u_a) 
				& \Rightarrow \mcal{A}^{(m)}_{Z_{1p}|u_a,\delta}
		\end{align}

		Applied to the $n$-symbols of the channel output we get:
		\begin{align}
			\rho^\prime_{u_1^n,u_2^n}  
				& =  \sum_{z_1^n} \ketbra{z_1^n}{z_1^n} \rho_{u_1^n,u_2^n} \ketbra{z_1^n}{z_1^n} \\
				& = \sum_{z_1^n}  p_{Z_{1p}^n|U_1^nU_2^n}(z_{1}^n|u_1^n,u_2^n) \ketbra{z_1^n}{z_1^n},
		\end{align}
		where $p_{Z_{1p}^n|U_1^nU_2^n}(z_{1}^n|u_1^n,u_2^n)$ is a product distribution
		built from the individual distributions $p_{Z_{1p}|U_1U_2}(z_{1}|u_a,u_b)$
		depending on the value of $u_{1i}$ and $u_{2i}$.
		Note also that the basis $\ket{z_1^n}$ is built from the different bases $\ket{z_{1i}^{(u_1)}}$
		according to whichever input symbol $u_{1i}$ is used.
		To make the above statements more explicit, we can permute order of the symbols in the 
		codeword so that they form contiguous blocks where the same input $u_a$ is used.
		
		\begin{align}
			\rho^\prime_{u_1^n,u_2^n} 
				& = \sum_{z_1^n}  p_{Z_{1p}^n|U_1^nU_2^n}(z_{1}^n|u_1^n,u_2^n) \ketbra{z_1^n}{z_1^n} \\
				& = \bigotimes_{u_a \in \mcal{U}_1 } 
					\left(  
						\prod_{j=1}^{m_1}
						p_{ Z_{1pj}| U_1U_2} ( z_{1j} |u_a,u_{2j} ) 
						\ketbra{ z_{1j} } { z_{1j}  } 
					\right),
		\end{align}
		where $m_a=N(u_a|u_1^n,u_2^n)$ and $\sum_a m_a = n$.
		In each $m_a$-dimensional block, the same basis is used for all symbols: $\ket{z_1^{(m_1)} } = \ket{ z_1^{(u_a)}}$.
		
		Using the pinching operator, we can reduce the lemma to  a question involving only classical probability distribution.
		Let $m_a=N(u_a|u_1^n)$ and decompose $\PIone$ into different blocks:
		\begin{align}
			\Tr\left[ \PIone \ \rhoNulul \right]  
				&= \Tr\left[ \bigotimes_{u_a} \left( \sum_{z_1^{(m_a)} \in \AtypGua } \ketbra{z_1^{(m_1)}}{z_1^{(m_1)}} \right) \ \rhoNulul \right] \\
				& = \prod_{u_a} \sum_{z_1^{(m_a)} \in  \AtypGua } p_{Z_{1p}^{(m_a)}|U^{(m_a)}_1U^{(m_a)}_2}(z_{1p}^{(m_a)}|u_a,u^{(m_a)}_2) \\
				& = \prod_{u_a}  \Pr\left\{
					   Z_{1p}^{(m_a)}|u_a,u^{(m_a)}_2 \in  \AtypGua
					   \right\} \\
				& \geq \prod_{u_a}  (1 - \eps'' ) \\
				& = 1 - |\mcal{U}_1|\eps'' = 1 - \eps'''.
		\end{align}
		
		We use a  similar argument as in the previous lemma.
		In each block we know that w.h.p.~$Z_{1p}^{(m_a)}|u_a,u^{(m_a)}_2 \in  \AtypGuaun$
		and $u_2^{(m_a)} \in \mcal{A}_{U_2|u_a}^{(m_a)}$ therefore it must be that
		$Z_{1p}^{(m_a)}u^{(m_a)}_2|u_a \in  \mcal{A}_{ZU_2|u_a}^{(m_a)}$.
		This in turn implies that 		
		$Z_{1p}^{(m_a)}|u_a \in \AtypGua$.
	\end{IEEEproof}

}{}

\bibliographystyle{IEEEtran}
\bibliography{interferenceChannel}

\begin{thebibliography}{10}
\providecommand{\url}[1]{#1}
\csname url@samestyle\endcsname
\providecommand{\newblock}{\relax}
\providecommand{\bibinfo}[2]{#2}
\providecommand{\BIBentrySTDinterwordspacing}{\spaceskip=0pt\relax}
\providecommand{\BIBentryALTinterwordstretchfactor}{4}
\providecommand{\BIBentryALTinterwordspacing}{\spaceskip=\fontdimen2\font plus
\BIBentryALTinterwordstretchfactor\fontdimen3\font minus
  \fontdimen4\font\relax}
\providecommand{\BIBforeignlanguage}[2]{{%
\expandafter\ifx\csname l@#1\endcsname\relax
\typeout{** WARNING: IEEEtran.bst: No hyphenation pattern has been}%
\typeout{** loaded for the language `#1'. Using the pattern for}%
\typeout{** the default language instead.}%
\else
\language=\csname l@#1\endcsname
\fi
#2}}
\providecommand{\BIBdecl}{\relax}
\BIBdecl

\bibitem{el2010lecture}
A.~El~Gamal and Y.-H. Kim, \emph{Network Information Theory}.\hskip 1em plus
  0.5em minus 0.4em\relax Cambridge University Press, 2012.

\bibitem{C72}
T.~M. Cover, ``Broadcast channels,'' \emph{IEEE Transacations on Information
  Theory}, vol.~18, no.~1, pp. 2--14, January 1972.

\bibitem{B73}
P.~P. Bergmans, ``Random coding theorem for broadcast channels with degraded
  components,'' \emph{IEEE Transactions on Information Theory}, vol.~19, no.~2,
  pp. 197--207, March 1973.

\bibitem{M79}
K.~Marton, ``A coding theorem for the discrete memoryless broadcast channel,''
  \emph{IEEE Transacations on Information Theory}, vol.~25, no.~3, pp.
  306--311, 1979.

\bibitem{wilde2011book}
M.~M. Wilde, \emph{From Classical to Quantum Shannon Theory}, 2011,
  arXiv:1106.1445.

\bibitem{H98}
A.~S. Holevo, ``The capacity of the quantum channel with general signal
  states,'' \emph{IEEE Transacations on Information Theory}, vol.~44, no.~1,
  pp. 269--273, 1998.

\bibitem{SW97}
B.~{Schumacher} and M.~D. {Westmoreland}, ``Sending classical information via
  noisy quantum channels,'' \emph{Physical Review A}, vol.~56, pp. 131--138,
  1997.

\bibitem{winter2001capacity}
A.~Winter, ``{The capacity of the quantum multiple-access channel},''
  \emph{IEEE Transacations on Information Theory}, vol.~47, no.~7, pp.
  3059--3065, 2001.

\bibitem{PhysRevA.63.032312}
\BIBentryALTinterwordspacing
A.~S. Holevo and R.~F. Werner, ``Evaluating capacities of bosonic gaussian
  channels,'' \emph{Physical Review A}, vol.~63, p. 032312, Feb 2001. [Online].
  Available: \url{http://link.aps.org/doi/10.1103/PhysRevA.63.032312}
\BIBentrySTDinterwordspacing

\bibitem{PhysRevLett.92.027902}
V.~Giovannetti, S.~Guha, S.~Lloyd, L.~Maccone, J.~H. Shapiro, and H.~P. Yuen,
  ``Classical capacity of the lossy bosonic channel: The exact solution,''
  \emph{Physical Review Letters}, vol.~92, no.~2, p. 027902, January 2004.

\bibitem{GHG13}
V.~Giovannetti, A.~S. Holevo, and R.~Garcia-Patron, ``A solution of the
  {Gaussian} optimizer conjecture,'' December 2013, arXiv:1312.2251.

\bibitem{YHD2006}
J.~Yard, P.~Hayden, and I.~Devetak, ``Quantum broadcast channels,'' \emph{IEEE
  Transacations on Information Theory}, vol.~57, no.~10, pp. 7147--7162,
  October 2011.

\bibitem{guha2007classical}
S.~Guha, J.~Shapiro, and B.~Erkmen, ``Classical capacity of bosonic broadcast
  communication and a minimum output entropy conjecture,'' \emph{Physical
  Review A}, vol.~76, no.~3, p. 032303, 2007.

\bibitem{DHL10}
F.~Dupuis, P.~Hayden, and K.~Li, ``A father protocol for quantum broadcast
  channels,'' \emph{IEEE Transacations on Information Theory}, vol.~56, no.~6,
  pp. 2946--2956, June 2010.

\bibitem{FHSSW11}
O.~Fawzi, P.~Hayden, I.~Savov, P.~Sen, and M.~M. Wilde, ``Classical
  communication over a quantum interference channel,'' \emph{IEEE Transactions
  on Information Theory}, vol.~58, no.~6, pp. 3670--3691, June 2012,
  arXiv:1102.2624.

\bibitem{S11a}
P.~Sen, ``Achieving the {Han-Kobayashi} inner bound for the quantum
  interference channel by sequential decoding,'' September 2011,
  arXiv:1109.0802.

\bibitem{RSW14}
J.~Radhakrishnan, P.~Sen, and N.~Warsi, ``One-shot {Marton} inner bound for
  classical-quantum broadcast channel,'' October 2014, arXiv:1410.3248.

\bibitem{SW12ISIT}
I.~Savov and M.~M. Wilde, ``Classical codes for quantum broadcast channels,''
  in \emph{Proceedings of the 2012 IEEE International Symposium on Information
  Theory}, Cambridge, MA, USA, 2012, pp. 721--725.

\bibitem{itit1999winter}
A.~Winter, ``Coding theorem and strong converse for quantum channels,''
  \emph{IEEE Transacations on Information Theory}, vol.~45, no.~7, pp.
  2481--2485, 1999.

\bibitem{hayashi2003general}
M.~Hayashi and H.~Nagaoka, ``General formulas for capacity of classical-quantum
  channels,'' \emph{IEEE Transacations on Information Theory}, vol.~49, no.~7,
  pp. 1753--1768, 2003.

\bibitem{GLM10}
V.~Giovannetti, S.~Lloyd, and L.~Maccone, ``Achieving the {Holevo} bound via
  sequential measurements,'' \emph{Physical Review A}, vol.~85, no.~1, p.
  012302, January 2012, arXiv:1012.0386.

\bibitem{eGvdM81}
A.~El~Gamal and E.~van~der Meulen, ``A proof of {Marton's} coding theorem for
  the discrete memoryless broadcast channel (corresp.),'' \emph{IEEE
  Transactions on Information Theory}, vol.~27, no.~1, pp. 120--122, January
  1981.

\bibitem{GA12}
A.~Gohari and V.~Anantharam, ``Evaluation of marton's inner bound for the
  general broadcast channel,'' \emph{IEEE Transactions on Information Theory},
  vol.~58, no.~2, pp. 608--619, February 2012.

\bibitem{JN10}
V.~Jog and C.~Nair, ``An information inequality for the bssc channel,''
  \emph{Proceedings of the UCSD Information Theory and Applications Workshop},
  2010.

\end{thebibliography}

\end{document}